\documentclass{amsart}
\usepackage{amsmath,amscd,amssymb,amsthm}

\newtheorem{mainthm}{Theorem}

\newtheorem{lemma}{Lemma}[section]

\newtheorem{prop}[lemma]{Proposition}
\newtheorem{thm}[lemma]{Theorem}
\newtheorem{cor}[lemma]{Corollary}

  \font\bib=cmbx8 \font\eightit=cmti8

\def\pz{\partial_z}
\def\pyi{\partial_{y_i}}
\def\pxi{\partial_{x_i}}
\def\oh{{\ts\frac{1}{2}}}

\def\cirm{C^\infty (\bR^m)}
\def\crm{\K (\bR^m)}
\def\prm{\Poly (\bR^m)}
\def\trm{\Tan (\bR^m)}
\def\vrm{\Vec (\bR^m)}

\def\cm{\K (M)}
\def\tm{\Tan (M)}
\def\vm{\Vec (M)}

\def\fs{\mathop{{\rm f}\sigma}\nolimits}
\def\ss{\mathop{{\rm s}\sigma}\nolimits}

\def\cal{\mathcal}

\def\D{{\cal D}} 
\def\E{{\cal E}} 
\def\F{{\cal F}}

\def\K{{\cal K}}
\def\L{{\cal L}}

\def\P{{\cal P}}

\def\S{{\cal S}}

\def\frak{\mathfrak}

\def\da{{\frak a}}

\def\db{{\frak b}}

\def\dc{{\frak c}}

\def\dg{{\frak g}}
\def\dgl{\dg\dl}
\def\dh{{\frak h}}

\def\dl{{\frak l}}

\def\dgo{{\frak o}}

\def\dgp{{\frak p}}

\def\ds{{\frak s}}
\def\dsl{\ds\dl}
\def\dsp{\ds\dgp}
\def\dt{{\frak t}}
\def\du{{\frak u}}
\def\dU{{\frak U}}

\def\dZ{{\frak Z}}

\def\Bbb{\mathbb}

\def\bC{\Bbb C}

\def\bN{\Bbb N}

\def\bP{\Bbb P}

\def\bR{\Bbb R}

\def\bZ{\Bbb Z}

\def\De{\Delta}

\def\l{\lambda}

\def\rmC{\mathop{\rm C}\nolimits}

\def\rms{\mathop{\rm s}\nolimits}
\def\rmT{\mathop{\rm T}\nolimits}

\def\la{\langle}

\def\ra{\rangle}

\def\Div{\mathop{\rm Div}\nolimits}
\def\DivC{\Div_{\rmC}}
\def\DivT{\Div_{\rmT}}
\def\End{\mathop{\rm End}\nolimits}
\def\gr{\mathop{\rm gr}\nolimits}
\def\FS{{\Sigma}}

\def\Hom{\mathop{\rm Hom}\nolimits}

\def\NQ{{\rm N}}
\def\oh{{\ts\frac{1}{2}}}

\def\Poly{\mathop{\rm Poly}\nolimits}
\def\poly{\mathop{\rm poly}\nolimits}
\def\PQ{{Q}}
\def\prm{\Poly(\bR^m)}

\def\rmp{\bR^m_{\poly}}
\def\scrm{\scriptsize\rm}

\def\Span{\mathop{\rm Span}\nolimits}
\def\SPQ{{\S Q}}

\def\Sym{\mathop{\rm Sym}\nolimits}
\def\Tan{\mathop{\rm Tan}\nolimits}
\def\thup{{\mbox{\scrm th}}}
\def\ts{\textstyle}
\def\Vec{\mathop{\rm Vect}\nolimits}
\def\vol{\omega}

\def\dog{differential operator}
\def\eg{{\em e.g.,\/}}

\def\ic{{infinitesimal character}}
\def\ie{{\em i.e.,\/}}
\def\iff{if and only if}
\def\irr{irreducible}

\def\lwv{lowest weight vector}
\def\meno{\medbreak\noindent}

\def\r{representation}

\def\tdm{tensor density module}

\def\th{\thinspace}
\def\uea{universal enveloping algebra}
\def\vf{vector field}

\title[Contact manifolds]{Linear differential operators on contact manifolds}

\author{Charles H.\ Conley}
\address{Department of Mathematics 
\\University of North Texas 
\\Denton TX 76203, USA} 
\email{conley@unt.edu}

\author{Valentin Ovsienko}
\address{CNRS, Institut Camille Jordan 
\\Universit\'e Claude Bernard Lyon~I
\\21 Avenue Claude Bernard, 69622 Villeurbanne Cedex, France}
\email{ovsienko@math.univ-lyon1.fr}

\begin{document}

\begin{abstract}
We consider differential operators between sections 
of arbitrary powers of the determinant line bundle over a contact manifold.
We extend the standard notions of the Heisenberg calculus:
noncommutative symbolic calculus, the principal symbol, and the contact order
to such differential operators.
Our first main result is an intrinsically defined ``subsymbol'' of a differential operator,
which is a differential invariant of degree one lower than that of the principal symbol.
In particular, this subsymbol associates a contact vector field 
to an arbitrary second order linear differential operator.
Our second main result is the construction of a
filtration that strengthens the well-known contact order filtration
of the Heisenberg calculus.
\end{abstract}


\maketitle

\section{Introduction}  \label{Intro} 

The space $\D(M)$ of linear differential operators 
on a smooth manifold $M$ has a rich geometric structure.
By the \textit{geometry} of $\D(M)$, 
we understand its structure as
a module over the group of all diffeomorphisms of $M$, 
and thereby also over its Lie algebra,
the space $\vm$ of smooth vector fields on $M$.
The most interesting geometric properties of $\D(M)$ are 
described by its invariants under the group of diffeomorphisms.

Additional structure on $M$ leads to a smaller group of diffeomorphisms,
and therefore a richer set of invariants of $\D(M)$.
Contact manifolds provide an important class of examples of geometric structures.
In this paper we study the geometric properties of $\D(M)$ viewed as a module over
the Lie algebra $\K(M)$ of all contact vector fields on $M$.
This viewpoint fits into the general framework of
\textit{Heisenberg calculus}, see \cite{BG88, EM98, vE10},
where the geometric structure is a codimension-1 distribution in $T(M)$.

Our first main result is the association of a contact vector field 
to an arbitrary second order linear differential operator in a contact-invariant manner.
Although we do not carry out the investigation here, this could provide
a means to associate topological invariants to second order operators.
We generalize the result to differential operators of arbitrary order, associating
to each a certain \textit{tensor density} on~$M$.
This tensor density is independent of the symbol of the operator
and may be thought of as a partial ``subsymbol''.

By a tensor density, we mean a section 
of a power of the determinant line bundle.
In fact, we state our results in the more general context 
of the spaces $\D_{\l,\mu}(M)$ of differential operators between such line bundles, 
rather than simply for differential operators on functions.
An interesting feature is appearance of
\textit{contact resonances},
that is, of special powers of the determinant line bundle
for which the geometric properties of differential operators are more complicated.
These resonances were already observed in \cite{FMP08}.
Let us mention that the usual case where $\l=\mu$ is non-resonant.

Our second main result is the existence of a filtration
refining the usual filtration given by the Heisenberg calculus.
Recall that differential operators on a contact manifold
have a \textit{contact order}, in which vector fields tangent to the
distribution are of order~1, and contact vector fields are of order~2.
We introduce a contact-invariant filtration on $\D_{\l,\mu}(M)$
for which, roughly speaking, tangential vector fields have order~1
and contact vector fields have order~3.
However, this filtration is not compatible with composition.
We prove its existence in the non-resonant case.

We remark that there is also an invariant double filtration
on the space of differential forms on~$M$ \cite{Ru94}.
In some ways, the situation for contact manifolds
appears to be analogous to that for foliated manifolds.
This may at first be surprising, 
as contact distributions are completely non-integrable,
but such analogies have been observed before \cite{ET98}.

Although our results and their applications are essentially geometric,
the proofs are algebraic.  The considerations are local, so we may work
in the Euclidean case, replacing $M$ by $\bR^m$, where $m=2\ell+1$.
Moreover, we need only consider the Lie algebra of
polynomial contact vector fields on $\bR^m$, which is the classical
infinite dimensional Cartan algebra $\K_m$; see \cite{Fu86}.

Our main theorems are proven using certain underlying structural results 
concerning the cohomology of $\K_m$ with coefficients in spaces of
\dog s between refined symbol modules.
These results in turn are obtained using a quantization map
which is equivariant with respect to the 
projective subalgebra $\ds_m$ of $\K_m$,
a maximal subalgebra isomorphic to $\dsp_{2(\ell+1)}$.
We prove the existence and uniqueness of this quantization map
using the description of the \ic s of $\ds_m$
given by the Harish-Chandra homomorphism.
We also calculate the map explicitly.

An $\ds_m$-module is said to have an \ic\ if the
center $\dZ(\ds_m)$ of the \uea\ of $\ds_m$ acts on it by scalars.
The \ic\ is then the resulting homomorphism from $\dZ(\ds_m)$ to $\bC$.
If an $\ds_m$-module has a finite 
Jordan-H\"older composition series of modules
with distinct \ic s, then the module splits as the direct sum of its
composition series modules.

Let us discuss at this point
the role of \ic s in other forms of quantization.
The {\em Casimir element\/} 
is the best known and simplest element of $\dZ(\ds_m)$.
It turns out that for the contact projective quantization 
studied in this paper, it is not sufficient
to consider the eigenvalues of the Casimir element alone,
because there are fine symbol modules
with distinct \ic s but identical Casimir eigenvalues.
This is in contrast with the situation for 
projective quantization with respect to
the full vector field Lie algebra $\vrm$, whose projective subalgebra is $\dsl_{m+1}$.
The full principal symbol modules have
\ic s under the action of $\dsl_{m+1}$, and these \ic s are distinct
\iff\ their Casimir eigenvalues are different \cite{Le00}.
Therefore in this setting there is no need to consider \ic s.

For conformal quantization, one replaces
the projective subalgebra with the conformal subalgebra
$\dgo_{p+1, q+1}$, a maximal subalgebra of $\Vec(\bR^{p+q})$.
As was first observed in \cite{DLO99}, in this setting
the Casimir element of $\dgo_{p+1, q+1}$ is not
sufficient to detect distinct \ic s among these submodules.
Complete results concerning the existence and uniqueness
of conformal quantization for \dog s between \tdm s 
have recently been obtained in \cite{Si09} and \cite{Mi11}.
It would be interesting to determine to what extent
\ic s can be used to replicate them.

The crucial property that allows us to apply algebraic results
in the geometric situation of an arbitrary contact manifold $M$ is the 
\textit{uniqueness} of the $\dsp_{2(\ell+1)}$-equivariant
quantization map.  For example, the subsymbol is first defined locally
in Darboux coordinates.
Its uniqueness then implies that it is defined globally on $M$.

This paper is organized as follows.
In Section~\ref{Results} we define the modules of tensor densities, \dog s,
and symbols, and formulate our main results.
In Section~\ref{K} we fix local Darboux coordinates and
review local properties of the contact Lie algebra.
In Section~\ref{S} we study the modules of Section~\ref{Results}
under the action of the projective subalgebra, using
\ic s to compute contact resonances.
In Section~\ref{PQ} we prove the existence and uniqueness of the
projective quantization, the natural projective equivalence
from symbols to \dog s.
Section~\ref{CRCs} contains the proofs of two of our main results:
the existence and uniqueness of the subsymbol and of the fine filtration.
Section~\ref{T} gives the explicit formula for the projective quantization
and the subsymbol and proves our third main result.

\section{Main results}  \label{Results}

Fix $m = 2\ell + 1$ odd, and let $M$ be a smooth $m$-dimensional manifold 
equipped with a contact distribution $\Xi$: 
a completely non-integrable distribution of codimension~1.  
As usual, locally we define the contact structure in terms 
of a contact form $\theta$ whose kernel is $\Xi$. 
The non-integrability  of $\Xi$ is equivalent to the fact that 
$\theta\wedge(d\theta)^\ell$ is a local volume form.
We define the subspace $\tm$ of $\vm$ to consist of the sections of $\Xi$,
that is, the vector fields annihilated by $\theta$.
We will refer to such vector fields as \textit{tangential vector fields.\/}

\subsection{Definitions} \label{Definitions}

We will use the following notation throughout this paper.
For $X \in \vm$, we write $L(X)$ for the associated \textit{Lie derivative.\/}
The non-negative integers will be denoted by $\bN$,
and the positive integers by $\bZ^+$.
For $x \in \bR$, we use the floor notation 
for the greatest integer~$\le x$
and the ceiling notation for the least integer~$\ge x$:
\begin{equation*}
   \lfloor x \rfloor := \sup \{ n \in \bZ: n \le x \}, \qquad
   \lceil x \rceil := \inf \{ n \in \bZ: n \ge x \}.
\end{equation*}

Within $\vm$ we have the Lie subalgebra $\cm$ of 
\textit{contact vector fields,} those which preserve~$\Xi$.
Contact vector fields are characterized locally as those whose 
Lie derivatives preserve the conformal class of $\theta$. 
More precisely, a vector field $X$ on $M$ is contact if
\begin{equation} \label{LXalpha}
   L(X)\theta = {\ts \frac{1}{\ell+1}}\, \Div(X) \theta,
\end{equation}
where $\Div$ is the divergence
with respect to the volume form $\theta \wedge (d \theta)^\ell$.
The complete non-integrability of $\Xi$ translates to
\begin{equation*}
   \vm = \cm \oplus \tm. 
\end{equation*}
This decomposition is invariant under the Lie action of $\cm$. 
Observe that $\cm$ is not invariant under multiplication by functions,
and $\tm$ is not a Lie algebra.  Let 
$$
\pi:\Vec(M)\to\K(M)
$$
be the projection along $\Tan(M)$.

We now make several definitions valid for arbitrary (not necessarily contact) manifolds.

\meno {\bf Definition.}
\begin{enumerate}
\item[(i)]
For $\l \in \bC$, let $|\Lambda^m T^*(M)|^\l$ be the line bundle of 
homogeneous functions of degree $\l$ on the determinant bundle.
The space $\F_\l(M)$ of\/ {\em tensor densities of degree~$\lambda$} 
consists of the smooth sections of 
$|\Lambda^m T^*(M)|^\l$ with \textit{complex coefficients.\/}
It is a module for $\vm$, and we write $L_\l(X)$
for the action of a vector field $X$ on it.

\item[(ii)]
Let $\D_{\l, \mu}(M)$ be the space of \dog s from $\F_\l(M)$ to $\F_\mu(M)$, 
and let $L_{\l, \mu}$ be the natural action of $\vm$ on it. 
For $k \in \bN$, let $\D^k_{\l, \mu}(M)$ be
the subspace of operators of order~$\le k$.
The spaces $\D^k_{\l,\mu}(M)$ comprise the \textit{order filtration}
of $\D_{\l,\mu}(M)$ and are invariant under $\vm$.

\item[(iii)]
We write $\delta$ for the difference between $\mu$ and $\l$:
\begin{equation*}
   \delta := \mu -\l.
\end{equation*}

\item[(iv)]
The space of \textit{principal symbols} of degree $k$ is the quotient 
\begin{equation*}
   \S^k_\delta(M)\ :=\ \D^k_{\l,\mu}(M)/\D^{k-1}_{\l,\mu}(M).
\end{equation*}
It is well-known that its $\Vec(M)$-module structure depends only on~$\delta$.

\item[(v)]
The \textit{principal symbol} is the natural projection
\begin{equation*} 
   \sigma^k_{\l,\mu}: \D^k_{\l,\mu}(M) \to \S^k_\delta(M).
\end{equation*}
\end{enumerate}

Let us give some natural examples of \tdm s.  
The simplest is $C^\infty(M)$, which is $\F_0(M)$.
In the contact setting, the following facts are well-known.

\begin{itemize}
\item
The adjoint action of the Lie algebra $\cm$ of 
contact \vf s on itself is equivalent to $\F_{-\frac{1}{\ell+1}}(M)$.
In other words, there is a $\K(M)$-equivalence
$$
X:\F_{-\frac{1}{\ell+1}}(M)\to\K(M),
$$
associating a contact vector field $X_\varphi$ to each element $\varphi$
of $\F_{-\frac{1}{\ell+1}}(M)$.
The tensor density $\varphi$ is called the
\textit{contact Hamiltonian} of $X_\varphi$.
The notion of the contact Hamiltonian is independent of the choice of
a contact form $\theta$.
However, fixing $\theta$ one can (locally) identify tensor densities and functions
and think of a contact Hamiltonian as of a function.

\item
The conformal class $C^\infty(M) \theta$ of the contact form $\theta$ is equivalent to 
$\F_{\frac{1}{\ell+1}}(M)$ as a $\cm$-module.  
\end{itemize}

In fact, the second statement follows from~(\ref{LXalpha}),
and the first follows from Lemma~\ref{Lagrange} below.
Thus the $\K(M)$-modules of contact Hamiltonians and contact forms are dual over $C^\infty(M)$.
We remark that the algebraic direct sum $\bigoplus_\l \F_\l(M)$ of all \tdm s
is a Poisson algebra under the Lagrange bracket.

The space $\D_{\l, \mu}(M)$ generalizes $\D_{0,0}(M)$,
which is the usual space of differential operators acting on functions.
Geometric properties of $\D_{\l, \mu}(M)$ vary with the parameters,
and the structure of $\D_{\l, \mu}(M)$ viewed as a $\K(M)$-module
can be special for certain values of $\l$ and $\mu$.
Let us stress the fact that differential operators between tensor densities
appear naturally in many geometric situations.
We mention for example the classical notion of the conformally invariant Laplace operator,
also known as the Yamabe Laplacian,
which is an element of 
$\D_{\frac{1}{2}-\frac{1}{m},\frac{1}{2}+\frac{1}{m}}(M)$.
The case $\l+\mu=1$ is particularly special.
This is the case where the notions of \textit{symmetric} and \textit{skew-symmetric}
operators are well-defined.
More generally, if $A\in\D_{\l, \mu}(M)$, then the adjoint operator $A^*$
belongs to $\D_{1-\mu,1- \l}(M)$.

\medskip

We now recall the classical notion of the Heisenberg order
of a differential operator on a contact manifold;
see for example \cite{vE10} and references therein.
 
\meno {\bf Definition.}

\begin{enumerate}

\item[(i)]
The space of differential operators of \textit{Heisenberg order} $\leq{}d$ is
$$
\P^d_{\l,\mu}(M)\ :=\ \Span\bigl\{
T_c \circ L_\l(Y_1) \circ \cdots \circ L_\l(Y_{t})\,:\,
T_c \in \D^c_{\l, \mu}(M),\, Y_i\in\Tan(M),\,
2c+t\leq{}d
\bigr\}.
$$
The spaces $\P^d_{\l,\mu}(M)$,
comprise the \textit{Heisenberg filtration} of $\D_{\l,\mu}(M)$.
They are invariant under $\K(M)$.

\item[(ii)]
The bifiltration $\D^{k,d}_{\l,\mu}(M):=\D^k_{\l,\mu}(M)\cap\P^d_{\l,\mu}(M)$
gives rise to the \textit{fine symbol} modules:
\begin{equation*}
   \FS^{k,d}_\delta(M)\ :=\ 
   \D^{k,d}_{\l,\mu}(M) / \bigl(\D^{k-1, d}_{\l,\mu}(M)+\D^{k,d-1}_{\l,\mu}(M)\bigr).
\end{equation*}

\item[(iii)]
The \textit{fine symbol} is the corresponding projection
\begin{equation*}
   \fs^{k,d}_{\l,\mu}: \D^{k,d}_{\l,\mu}(M) \to \FS^{k,d}_\delta(M).
\end{equation*}

\end{enumerate}

\medskip

In the simplest case $k=1$ and $\delta=0$, 
$\FS^{1,1}_0(M)$ is $\Tan(M)$ and $\FS^{1,2}_0(M)$ is $\K(M)$.
Here $ \fs^{1,2}_{\l,\l}$ is nothing but the projection $\pi$
defined above.  More generally, 
it follows from Proposition~\ref{LXf} below that
\begin{equation*} 
   \FS^{k,2k}_\delta(M) \cong \F_{\delta-\frac{k}{\ell+1}}(M).
\end{equation*}
Therefore $\fs^{k,2k}_{\l,\mu}$ may be regarded as 
a $\K(M)$-equivariant linear projection from 
$\D^k_{\l,\mu}(M)$ to\/ $\F_{\delta-\frac{k}{\ell+1}}(M)$.
This type of equivariant assignment of a tensor density to a \dog\
is known as a \textit{differential invariant.\/}

\meno {\bf Definition.}
We say that $\delta$ is {\em contact-resonant\/} if it lies in the set
\begin{equation*}
\left\{\frac{1}{\ell+1} + \frac{n}{2(\ell+1)}
\,\Big|\,n\in\bN \right\}.
\end{equation*}

\medskip

We will  see in Section~\ref{ICS} that contact resonances arise from the representation theory
of $\dsp_{2(\ell+1)}$.

\subsection{The subsymbol}

Our first main theorem gives a new contact differential invariant.

\begin{mainthm} \label{subsymbol}
If $\delta$ is not contact-resonant, then there exists
a unique\/ $\K(M)$-equivariant linear map
\begin{equation*}
   \ss^k_{\l,\mu}: \D^k_{\l,\mu}(M) \to \FS^{k-1,\,2(k-1)}_\delta(M)
\end{equation*}
whose restriction to\/ $\D^{k-1}_{\lambda,\mu}(M)$
is\/ $\fs^{k-1,\,2(k-1)}_{\lambda,\mu}$.
\end{mainthm}

We refer to $\ss^k_{\l,\mu}$ as the\/ {\em contact subsymbol.}
We will give an explicit formula for it in Proposition~\ref{subsymbol formula}.
It may be regarded
as a $\K(M)$-equivariant projection from $\D^k_{\lambda,\mu}(M)$
to $\F_{\delta - \frac{k-1}{\ell+1}}(M)$.

We remark that in the general self-adjoint case, where $\lambda + \mu = 1$
and $k$ is arbitrary, the existence of such a differential invariant is obvious.  
Indeed, for  $T$ in $\D^k_{\lambda,\mu}(M)$, the operator $T - (-1)^k T^*$
is in $\D^{k-1}_{\lambda,\mu}(M)$, and so can be projected to
$\FS^{k-1,\,2(k-1)}_\delta(M)$.

Since $\F_{-\frac{1}{\ell+1}}(M)$ is equivalent to $\cm$,
the case that $k=2$ and $\mu=\l$ is of particular interest,
as there the differential invariant given by the contact subsymbol may be viewed as
a contact vector field.  
In other words, for all $\l\in\bC$, the subsymbol $\ss^2_{\lambda,\lambda}$
defines a\/ $\cm$-equivariant projection from
$\D^2_{\lambda,\lambda}(M)$ to $\K(M)$.

In order to give an intrinsically defined and manifestly contact-invariant formula
for $\ss^2_{\lambda,\lambda}$, observe that any second order differential
operator can be represented as a linear combination of compositions
of vector fields.
On contact manifolds, contact vector fields and tangential vector fields
are intrinsically distinguished.
Thus we are led to express an arbitrary second order operator on $\F_\l(M)$
as a linear combination of operators of the form
\begin{equation}
\label{SecOp}
\begin{array}{rcl}
T&=&L_{\l}(X_{\varphi_1})\circ{}L_{\l}(X_{\varphi_2})+L_{\l}(X_{\varphi_3})\circ{}L_{\l}(Y_1)+
L_{\l}(Y_2)\circ{}L_{\l}(Y_3)\\[8pt]
&&+L_{\l}(X_{\varphi_4})+L_{\l}(Y_4)+f,
\end{array}
\end{equation}
where the $\varphi_i$ are arbitrary contact Hamiltonians,
the $Y_i$ are tangential vector fields, and $f$ is a function.

\begin{mainthm} 
\label{subsOrd2}
The subsymbol\/ $\ss^2_{\lambda,\lambda}(T)$ is the contact vector field
$$
\ts
\oh\bigl[X_{\varphi_1},X_{\varphi_2}\bigr]-
\bigl(\frac{\ell+1}{\ell+2}\bigr)\bigl(\l-\oh\bigr)X_{L(Y_1)\varphi_3}+
\oh\pi\bigl[Y_2,Y_3\bigr]+X_{\varphi_4},
$$
where\/ $L(Y_1)\varphi_3$ denotes the natural action of\/ $Y_1$ on the
$-\frac{1}{\ell+1}$-density\/ $\varphi_3$.
\end{mainthm}

Let us comment on this formula.
It only contains natural operations, so it is clearly contact-invariant.
Conversely, equivariance with respect to $\K(M)$ (in fact the affine subalgebra suffices)
implies that $\ss^2_{\lambda,\lambda}(T)$ has to be of the form
$$
\ts
c_{12}\bigl[X_{\varphi_1},X_{\varphi_2}\bigr]+
c_{13}X_{L(Y_1)\varphi_3}+
c_{23}\pi\bigl[Y_2,Y_3\bigr]+c_4X_{\varphi_4},
$$
where the $c$'s are constants.
The normalization condition on $\D^1_{\l,\l}$ gives $c_4=1$.
Skew-symmetrizing the expression then yields $c_{12}=c_{23}=\oh$.
Symmetrizing the expression implies that $c_{13}$ vanishes in the self-adjoint case $\l=\oh$,
but its exact form must be deduced by computation.

The main content of the theorem is that the formula is actually well-defined.
Indeed, the choice of the $\varphi_i$ and $Y_i$ in~(\ref{SecOp}) is not unique:
one can write an operator as a linear combination of such expressions in many
different ways.  However, the formula is independent of the choice.
Moreover, the uniqueness statement of Theorem~\ref{subsymbol} implies that, up to a scalar,
this is not true for any other choice of the $c$'s.

\subsection{The fine filtration}

In order to explain the significance of our next theorem, 
consider the following arrangement of the fine symbol modules
(we have omitted $M$ and $\delta$ for clarity):

\begin{equation*} 
\begin{array}{llllllllll}
   & & & & & & & & \FS^{6,12}&\\[10pt]
   & & & & & && \FS^{5,10}\ & \FS^{6,11}&\\[10pt]
   & & & & & \FS^{4,8}& & \FS^{5,9}\ & \FS^{6,10}&\\[10pt]
   & & & & \FS^{3,6}\ & \FS^{4,7}&& \FS^{5,8}\ & \FS^{6,9}\ & \cdots \\[4pt]
   \cline{3-3}\\[-8pt]
   & & \FS^{2,4}&\!\vline\!\!\!\! & \FS^{3,5}\ 
   & \FS^{4,6} & & \FS^{5,7}\ & \FS^{6,8}&\\[4pt]
   \cline{5-6}\\[-8pt]
   & \FS^{1,2}\ & \FS^{2,3}&& \FS^{3,4}\ & \FS^{4,5}
   &\!\vline & \FS^{5,6}\ & \FS^{6,7}& \\[4pt]
   \cline{8-9}\\[-8pt]
   \FS^{0,0}\ & \FS^{1,1}\ & \FS^{2,2}& & \FS^{3,3}\ 
   & \FS^{4,4}& & \FS^{5,5}\ & \FS^{6,6}&\!\!\vline 
\end{array} \end{equation*} 
\bigskip

\noindent
Observe that the graded module of $\S^k_\delta(M)$ 
defined by the bifiltration $\D^{k,d}_{\l,\mu}(M)$ is the ``vertical'' sum
$$
\gr\S^k_\delta(M)=\bigoplus_{k \le d \le 2k}\FS^{k,d}_\delta(M).
$$
The graded module of $\P^d_{\l,\mu}(M)/\P^{d-1}_{\l,\mu}(M)$ is the
``slope $-1$'' sum
$$
\gr\bigl(\P^d_{\l,\mu}(M)/\P^{d-1}_{\l,\mu}(M)\bigr)=
\bigoplus_{\lceil\frac{d}{2}\rceil \le k \le d}\FS^{k,d}_\delta(M).
$$

The content of our next theorem is that there exists
a $\K(M)$-invariant filtration that strengthens the filtration $\P^d_{\l,\mu}(M)$.
The graded modules of its subquotients are the  ``slope~$-\frac{1}{2}$'' sums.

\begin{mainthm} \label{bifilt}
Assume that $\delta$ is not contact-resonant.
Then there is a unique\/ $\K(M)$-invariant filtration of\/ $\D_{\l,\mu}(M)$,
\begin{equation*}
\D^{(0)}_{\l,\mu}(M)
\subset\cdots\subset
\D^{(b)}_{\l,\mu}(M)\subset
\D^{(b+1)}_{\l,\mu}(M)
\subset\cdots,
\end{equation*}
such that the graded module of\/ $\D^{(b)}_{\l,\mu}(M)$ is given by
\begin{equation*}
\mathrm{gr}\,\D^{(b)}_{\l,\mu}(M)=
\bigoplus_{2d-k\leq{}b}\,\FS^{k,d}_\delta(M).
\end{equation*}
\end{mainthm}

\medskip
For example,
$\mathrm{gr}\bigl(\D^{(6)}_{\l,\mu}(M)/\D^{(5)}_{\l,\mu}(M)\bigr)
=
\FS^{6,6}_\delta(M)\oplus{}\FS^{4,5}_\delta(M)\oplus{}\FS^{2,4}_\delta(M),
$
as indicated by the boundaries in the diagram above.
We will define $\D^{(b)}_{\l,\mu}(M)$ via the projective quantization:
see Section~\ref{C proof}.

\subsection{Comments on non-existence and conjectures}

In this paper we are concerned with existence results
rather than non-existence results, but we remark that
in the contact-resonant case
our theorems are false for most values of $\l$.
For example, if $k = \delta (\ell + 1) - \ell$
in Theorem~\ref{subsymbol}, then there exists no subsymbol
except in the self-adjoint case $\lambda + \mu = 1$,
where there exists a 1-parameter family of such maps.
Non-existence results can be interpreted in terms of cohomological 
obstructions.  We conjecture that the filtration of Theorem~\ref{bifilt}
does not exist in the contact-resonant case.
These questions will be addressed elsewhere.

A module is called \textit{uniserial,\/} 
or \textit{completely indecomposable,\/}
if it has a unique maximal invariant filtration.  
Such modules may be thought of as opposite to completely reducible modules.
We conjecture that for generic values of $\l$ and $\mu$,
$\D^{(b)}_{\l,\mu}(M) / \D^{(b-1)}_{\l,\mu}(M)$ is uniserial,
because we expect that in the Euclidean case, there is a non-trivial projectively relative
1-cohomology class of $\K_m$ linking $\FS^{k,d}_\delta$ to $\FS^{k-2,d+1}_\delta$:
an analog of the Schwarzian derivative.
This conjecture says essentially that
the bifiltration $\D^k_{\l,\mu}(M)\cap\D^{(b)}_{\l,\mu}(M)$ is the best possible.  
In other words, if the conjecture is true then there 
is no $\K(M)$-invariant filtration of $\D_{\l,\mu}(M)$ 
whose elements are composed of all those fine symbol modules 
on or below the lines of any fixed slope shallower than~$-\frac{1}{2}$
passing through the above diagram.

\section{The Euclidean contact Lie algebra} \label{K}

Since all of the theorems in Section~\ref{Results} are local,
their proofs essentially reduce to the case $M = \bR^m$.  
Therefore in this section we establish notation 
and state some well-known results for Euclidean contact manifolds:
the proofs are straightforward and are usually omitted.  
All $\cirm$-modules of finite rank are equipped 
with their usual topologies as Frechet spaces, 
and by definition all $\Hom$ spaces between such modules 
include only continuous linear maps.  

\subsection{Darboux coordinates}\label{DarbSect}

Fix coordinates $x_i$, $y_i$, and $z$ on $\bR^m$,
where $m=2\ell+1$ and $1 \le i \le \ell$.  
Henceforth we use {\em Einstein's summation convention:\/} unless stated otherwise, 
{\em repeated indices are summed over from~$1$ to~$\ell$.\/}  
Let $\theta$ and $\vol$ be the standard contact and volume forms on $\bR^m$: 
\begin{equation*}
   \theta := dz + \oh (x_i dy_i - y_i dx_i), 
   \qquad
   \vol := {\ts \frac{1}{\ell!}} \th \theta \wedge (d \theta)^\ell
   = dz \wedge {\ts \bigwedge_1^\ell} (dx_i \wedge dy_i). 
\end{equation*}
The standard divergence operator $\Div: \vrm \to \cirm$ is defined by
\begin{equation*}
   L(X)\, \vol = \Div(X) \vol.
\end{equation*}

Recall from Section~\ref{Results} the
definitions of the Lie algebra of contact vector fields and
the space of tangent vector fields:
\begin{eqnarray*}
   \crm &:=& \bigl\{ X \in \vrm: 
   L(X)\, \theta = {\ts \frac{1}{\ell + 1}} \Div(X) \theta \bigr\}, \\[6pt]
   \trm &:=& \bigl\{ X \in \vrm: \la \theta, X \ra = 0 \bigr\}.
\end{eqnarray*}
The space $\trm$ is a module over $\crm$, and so one has
the $\crm$-invariant decomposition
\begin{equation*}
   \vrm = \crm \oplus \trm. 
\end{equation*}
It is important to keep in mind that while $\trm$ is closed 
under multiplication by smooth functions, $\crm$ is not.  
We now give explicit descriptions of both spaces.

Define the following {\em Euler operators:\/}
\begin{equation*}
   \E_z := z \pz, \qquad \E_{xy} := x_i \pxi + y_i \pyi.
\end{equation*}
For $1 \le i \le \ell$, define the vector fields
\begin{equation*}
   A_i := \pxi + \oh y_i \pz, \qquad B_i := -\pyi + \oh x_i \pz.
\end{equation*}
The following statements are classical.

\begin{itemize}

\item
There is a linear bijection\/ $X: \cirm \to \crm$ mapping $f$ to $X_f$, 
the unique contact vector field such that\/ $\la \theta, X_f \ra = f$.  
It has the following explicit formulas:
\begin{equation} \label{Xf}
\begin{array}{rcl}
   X_f &=& f\pz + B_i(f) A_i - A_i(f) B_i \\[6pt]
   &=& \bigl( 1 - \oh \E_{xy} \bigr)(f) \pz
   + \oh \pz(f) \E_{xy} + \bigl(\pxi(f) \pyi - \pyi(f) \pxi\bigr).
\end{array}
\end{equation}

\item
$\{ A_i, \th B_i: 1 \le i \le \ell \}$ 
is a basis of\/ $\trm$ over\/ $\cirm$, and
\begin{equation*}
   [A_i, B_j] = \delta_{ij} \pz, 
   \qquad [\pz, A_i] = 0,
   \qquad [\pz, B_i] = 0.
\end{equation*}
\end{itemize}

We remark that one can verify~(\ref{LXalpha}) 
directly in this setting by checking that
\begin{equation} \label{explicit LXalpha}
   L(X_f) \,\theta = \pz(f)\,\theta, \qquad \Div(X_f) = (\ell + 1) \pz(f) .
\end{equation}

\meno {\bf Definition.}
The {\em Lagrange bracket\/} $\{f,g\}$ on $\cirm$ is defined by 
\begin{equation*}
   X_{\{f, g\}} := [X_f, X_g].
\end{equation*}
The following formulas for $\{f, g\}$ may be deduced from~(\ref{Xf}).

\begin{lemma} \label{Lagrange}
The Lagrange bracket is given by
\begin{eqnarray*}
   \{f, g\} &=& X_f(g) - g \pz(f) \ =\ f \pz(g) - X_g(f) \\[6pt]
   &=& f \pz(g) - \pz(f) g + B_i(f) A_i(g) - A_i(f) B_i(g) \\[6pt]
   &=& \bigl( 1 - \oh \E_{xy} \bigr)(f) \pz(g) 
   - \bigl( 1 - \oh \E_{xy} \bigr)(g) \pz(f)
   + \bigl(\pxi(f) \pyi(g) - \pyi(f) \pxi(g)\bigr).
\end{eqnarray*}
\end{lemma}

\subsection{Tensor density modules} \label{TDMs}

Recall from Section~\ref{Results} 
the tensor density module $\F_\lambda(\bR^m)$ of $\vrm$.
As a vector space, $\F_\lambda(\bR^m)$ is nothing but $\cirm$.
However, the action of $\vrm$ depends on $\lambda$ and is given by
\begin{equation*}
   L_\lambda(X) (g) := 
  X(g) + \lambda \Div(X)\,g,
\end{equation*}
where $g\in\cirm$ and $\lambda\in\mathbb{R}$.
Since the volume form $\vol$ is global, 
we may regard $\F_\lambda(\bR^m)$ as $\vol^\lambda \cirm$,  
so that the action of $\vrm$ is identified with the usual Lie derivative:
\begin{equation*}
   L(X) (\vol^\lambda g) = \vol^\lambda \,L_\lambda(X) (g).
\end{equation*}
We remark that the full family 
$\bigl\{ \F_\lambda(\bR^m): \lambda \in \bC \bigr\}$
of $\vrm$-modules can be understood algebraically as
a non-trivial deformation of the module $\cirm$.

We will consider $\F_\lambda(\bR^m)$ as a module over
the subalgebra of contact vector fields $\crm$.
In light of~(\ref{LXalpha}) and the global contact form $\theta$, 
we may regard $\F_\lambda(\bR^m)$ in this context
as either $\vol^\lambda \cirm$ or $\theta^{\lambda (\ell+1)} \cirm$.  
In particular, (\ref{explicit LXalpha}) gives
\begin{equation*}
   L(X_f) (\theta^{\lambda (\ell + 1)} g)
   = \theta^{\lambda (\ell + 1)} L_\lambda(X_f) (g)
   = \theta^{\lambda (\ell + 1)} \bigl(X_f(g)
   + \lambda (\ell + 1) \pz(f) g \bigr).
\end{equation*}

As mentioned in Section~\ref{Results}, the adjoint action of $\crm$ on itself
is equivalent to the module of $-\frac{1}{\ell + 1}$-densities.
The following definition and lemma state this formally.

\meno {\bf Definition.}
Henceforth we regard $X$ as the map 
\begin{equation*}
   X: \F_{-\frac{1}{\ell + 1}}(\bR^m) \to \crm, 
   \qquad X(\theta^{-1} f) := X_f.
\end{equation*}

\begin{lemma} \label{X}
The map\/ $X$ is a linear bijection and a\/ $\crm$-equivalence.
\end{lemma}

\meno{\em Proof.\/}
By Lemma~\ref{Lagrange}, $\{f,g\}=L_{-\frac{1}{\ell + 1}}(X_f) (g)$.
Therefore $X$ intertwines the $\crm$-action on 
$\F_{-\frac{1}{\ell + 1}}(\bR^m)$ and the adjoint action.  $\Box$

\subsection{Differential operator modules} \label{DOMs}

We now turn to the focus of the paper, the modules $\D_{\l,\mu}(\bR^m)$ 
of \dog s from $\F_\l(\bR^m)$ to $\F_\mu(\bR^m)$.  
We can write the action $L_{\l,\mu}$ of $\vrm$ on these modules 
concretely as follows:
\begin{equation*}
   L_{\lambda, \mu}(X) (T) := L_\mu(X) \circ T - T \circ L_\lambda(X).
\end{equation*}

The structure of the spaces $\D_{\l,\mu}(\bR^m)$
viewed as $\vrm$-modules has been thoroughly studied;
see for example \cite{LMT96, DO97, GMO05,Co09} and the references therein.
We will be interested in these spaces viewed as $\crm$-modules,
as which they are less rigid, because $\crm$ is smaller than $\vrm$.
In particular, the $\D_{\l, \mu}(\bR^m)$ admit more 
$\crm$-invariant operations than they do $\vrm$-invariant operations,
such as projections to tensor fields.
We will understand such operations as differential invariants.

\subsection{Symbol modules} \label{SMs}

In Section~\ref{Results} we defined the principal symbol modules $\S^k_\delta(\bR^m)$
and the fine symbol modules $\FS^{k,d}_\delta(\bR^m)$, $k \le d \le 2k$.
We will write $\S_\delta(\bR^m)$ and $\FS_\delta(\bR^m)$
for the {\em total symbol module\/} 
and the {\em total fine symbol module,\/} respectively; 
the graded modules associated to the order filtration 
and the bifiltration of $\D_{\l,\mu}(\bR^m)$:
\begin{equation*}
   \S_\delta(\bR^m)\ :=\ 
   \bigoplus_{k=0}^\infty \S^k_\delta(\bR^m), \qquad
   \FS_\delta(\bR^m)\ :=\ 
   \bigoplus_{k=0}^\infty  \bigoplus_{d=k}^{2k}\FS^{k,d}_\delta(\bR^m).
\end{equation*}

Let $L^\S_{\delta}$ be the natural action 
of $\vrm$ on $\S_\delta(\bR^m)$,
and let $L^{\FS}_\delta$ be the natural action 
of $\crm$ on $\FS_\delta(\bR^m)$.
Our next proposition gives formulas for 
$L^\S_\delta(X_f)$ and $L^{\FS}_\delta(X_f)$.
In order to state it we must develop 
a variation of the usual symbol calculus 
which is adapted to the fine filtration.  
Let $\alpha_i$, $\beta_i$, and $\zeta$ be the symbols 
associated to the vector fields $A_i$, $B_i$, and $\pz$.  
More explicitly,
\begin{equation}
\label{abxi}
\alpha_i=\xi_{x_i}+\oh\,y_i\xi_z,
\qquad
\beta_i=-\xi_{y_i}+\oh\,x_i\xi_z,
\qquad
\zeta=\xi_z,
\end{equation}
where $\xi_{x_1}, \xi_{y_1}, \ldots, \xi_{x_\ell}, \xi_{y_\ell}, \xi_z$
are the coordinates on $T^*\bR^{m}$ dual to $x_1, y_1, \ldots, x_\ell, y_\ell, z$.

We shall abuse notation and use $\alpha_i$ and $\beta_i$
also to denote fine symbols.  Thus
\begin{equation} \label{symbol bases} \begin{array}{rcl}
   \S^k_\delta (\bR^m) &=& \Span_{\cirm}\bigl\{ 
   \zeta^c \alpha^I \beta^J: |I| + |J| + c = k \bigr\}, \\[6pt]
   \FS^{k,d}_\delta (\bR^m) &=& \Span_{\cirm}\bigl\{ 
   \zeta^{d-k} \alpha^I \beta^J: |I| + |J| = 2k - d \bigr\},
\end{array} \end{equation}
where $I$ and $J$ are multi-indices: $I=(I_1,\ldots,I_\ell)$ and $J=(J_1,\ldots,J_\ell)$.
Note that in local coordinates we do not write the shift $\omega^\delta$
in the tensor density degree explicitly.

Any \dog\ from one symbol module to another, 
or from one fine symbol module to another, 
may be written as a linear combination
over $\cirm$ of products of the operators
\begin{equation*}
   A_i,\ B_i,\ \pz,\ \alpha_i,\ \beta_i,\ \zeta,\ 
   \partial_{\alpha_i},\ \partial_{\beta_i},\ \partial_\zeta.
\end{equation*}
Such a combination is to be interpreted as follows.  
The operators $A_i$, $B_i$, and $\pz$ act solely 
on the $\cirm$ coefficients of the basis elements in~(\ref{symbol bases}), 
while the remaining operators act solely on the basis elements themselves.  

Imitating the definitions of $\E_z$ and $\E_{xy}$, we set
\begin{equation*}
   \E_\zeta := \zeta \partial_\zeta, \qquad \E_{\alpha \beta} := 
   \alpha_i \partial_{\alpha_i} + \beta_i \partial_{\beta_i}.
\end{equation*}

\begin{prop} \label{LXf}

\begin{enumerate}

\item[(i)]
The action $L_\delta^{\FS}$ of\/ $\crm$ on\/ $\FS_\delta(\bR^m)$ is
\begin{equation*} 
\begin{array}{rcl}
   L_\delta^{\FS}(X_f) &=& f \pz + \th B_i(f) A_i - \th A_i(f) B_i + \th \pz(f)
   \bigl(\delta (\ell + 1) - \E_\zeta - \oh \E_{\alpha \beta} \bigr)
   \\[6pt]
   &&+ \th \oh (A_i B_i + B_i A_i)(f) (\beta_i \partial_{\beta_j} 
   - \alpha_i \partial_{\alpha_j})
   + \th A_i A_j(f) \beta_i \partial_{\alpha_j}
   - \th B_i B_j(f) \alpha_i \partial_{\beta_j}.
\end{array} 
\end{equation*}

\item[(ii)]
The action $L_\delta^{\FS}$ of\/ $\crm$ on\/ $\S_\delta(\bR^m)$ is
\begin{equation*} 
   L_\delta^{\S}(X_f) =
   L_\delta^{\FS}(X_f)+\bigl(\pz A_i(f) \beta_i 
   - \pz B_i(f) \alpha_i \bigr) \partial_\zeta.
\end{equation*}

\end{enumerate}
\end{prop}

\meno{\em Proof.\/}
Taking in to account the shift $\omega^\delta$ in tensor density degree,
$L^\S$ and $L^{\FS}$ are derivations in an obvious sense.
Therefore it is only necessary to check the formulas on the generators 
$\alpha_i$, $\beta_i$, and $\zeta$ and on functions $g$.  
Keep in mind that $A_i$ and $B_i$ do not commute, 
although $\alpha_i$ and $\beta_i$, being symbols, do.
$\Box$

\medbreak

The difference between $L_\delta^{\S}$ and $L_\delta^{\FS}$ is due to the fact that
$\FS_\delta(\bR^m)$ is the graded module of $\S_\delta(\bR^m)$.
Observe that $L_\delta^{\S}-L_\delta^{\FS}$ maps $\FS^{k,d}_\delta(\bR^m)$ to 
$\FS^{k,d-1}_\delta(\bR^m)$.

Henceforth we will frequently drop the argument $\bR^m$ 
of the various tensor density, \dog, and symbol modules,
writing simply $\F_\lambda$, $\D_{\lambda, \mu}$, 
$\S^k_\delta$, $\FS^{k,d}_\delta$, and so on.

\section{The projective subalgebra} \label{S}

Here we recall the {\em projective subalgebra\/} 
$\ds_m$ of $\crm$, which is isomorphic to $\dsp_{2(\ell+1)}$.
Restriction of $\crm$-modules to $\ds_m$ will be central
to our strategy throughout this paper, for two reasons.
First, $\ds_m$ is a maximal polynomial subalgebra of $\crm$,
and it turns out that for most values of their parameters,
the \tdm s and the fine symbol modules are not only 
algebraically \irr\ under $\crm$, they remain so under restriction to $\ds_m$.
(By ``algebraically \irr'', we mean \irr\ in the polynomial category.)
Second, $\ds_m$ is a finite dimensional semisimple Lie algebra,
and so we can bring the \r\ theory of such algebras to bear
on the restricted modules.

In fact, the restrictions to $\ds_m$ of the \tdm s and fine symbol modules
are duals of $\dsp_{2\ell}$-relative Verma modules.
Our approach to the construction of the projective quantization
referred to in the introduction will be to observe that for generic $\delta$,
the fine symbol modules composing $\D_{\l,\mu}$ have distinct
infinitesimal characters under the action of $\ds_m$.
This implies that there is a unique $\ds_m$-equivariant splitting
of $\D_{\l,\mu}$ into the sum of its fine symbol modules.
This splitting may be regarded as a projectively invariant total symbol.
The projective quantization is, by definition, its inverse.

In addition to the projective subalgebra $\ds_m$, 
two other subalgebras of $\crm$ will be important to us:
the {\em affine subalgebra\/} $\dt_m$ and its nilradical $\du_m$.

\subsection{The projective subalgebra of the full vector field Lie algebra} \label{SlCoord}

We first recall the definitions of the analogous subalgebras
of the full vector field Lie algebra.  Let $u_1, \ldots, u_m$
be any coordinates on $\bR^m$.  The full Euler operator is
$$
\E_u:=\sum_{i=1}^m u_i \partial_{u_i}.
$$ 
Within $\Vec(\bR^m)$ we have the projective subalgebra $\da_m$, 
the affine subalgebra $\db_m$, 
and the constant coefficient subalgebra $\dc_m$:
\begin{equation*} \begin{array}{rcl}
   \da_m &:=& \Span_\bC \bigl\{ \partial_{u_i}, 
   u_j \partial_{u_i}, u_j \E_u: 1 \le i, j \le m \bigr\}, \\[6pt]
   \db_m &:=& \Span_\bC \bigl\{ \partial_{u_i}, 
   u_j \partial_{u_i}: 1 \le i, j \le m \bigr\}, \\[6pt]
   \dc_m &:=& \Span_\bC \bigl\{ \partial_{u_i}: 
   1 \le i \le m \bigr\}.
\end{array} \end{equation*}

Clearly $\da_m \supset \db_m \supset \dc_m$.  
In fact, $\da_m$ is isomorphic to $\dsl_{m+1}$, and
$\db_m$ is a maximal parabolic subalgebra of $\da_m$
with Levi factor $\dgl_m$ and nilradical $\dc_m$.
The center of the Levi factor is $\bC \E_u$.

There is a standard conceptual proof 
of $\da_m \cong \dsl_{m+1}$ which we briefly sketch.  
Let $u_0, \ldots, u_m$ be coordinates on $\bR^{m+1}$, 
and note that $\Span_\bC \{ u_j \partial_{u_i}: 0 \le i, j \le m \}$ 
is a subalgebra of $\Vec (\bR^{m+1})$ isomorphic to $\dgl_{m+1}$.  
Restricting the action of $\Vec (\bR^{m+1})$ to functions on $\bP^m$ defines 
the canonical projection from this copy of $\dgl_{m+1}$ to $\dsl_{m+1}$.  
Regarding $\bP^m$ locally as the hyperplane defined by $u_0 = 1$ 
and identifying this hyperplane with $\bR^m$
yields the isomorphism.

\subsection{The projective subalgebra of the contact Lie algebra}

We now define the projective and affine subalgebras of $\crm$: 
they are simply the intersections of 
the projective and affine subalgebras of $\vrm$ with $\crm$.

\meno{\bf Definition.}
The\/ {\em projective subalgebra\/} $\ds_m$ 
and the\/ {\em affine subalgebra\/} $\dt_m$
of\/ $\crm$ are
\begin{equation*}
   \ds_m := \da_m \cap \crm, \qquad
   \dt_m := \db_m \cap \crm.
\end{equation*}
We define also the following subspaces of $\crm$:
\begin{eqnarray*}
   \du_m &:=& \Span_\bC \bigl\{ X_1, X_{x_i}, X_{y_i}: 
   1\le i \le \ell \bigr\}, \\[6pt]
   \dl_m^{\rms} &:=& \Span_\bC \bigl\{ X_{x_i y_j}, X_{x_i x_j}, X_{y_i y_j}: 
   1\le i, j \le \ell \bigr\}, \\[6pt]
   \dl_m &:=& \dl_m^{\rms} \oplus \bC X_z.
\end{eqnarray*}

\medbreak
Observe that $X_z = \E_z + \oh \E_{xy}$.
This is the natural Euler operator in $\crm$.
The space $\dl_m$ is the $0$-weight space of its adjoint action,
and $\du_m$ is the sum of its $-\oh$-eigenspace 
$\Span_\bC \{X_{x_i},\th X_{y_i}:\th 1 \le i \le \ell \}$ 
and its $-1$-eigenspace $\bC X_1$.

Next we give explicit descriptions of $\ds_m$ and $\dt_m$
and prove that $\ds_m$ is symplectic.
We also show that $\du_m$, $\dl_m^{\rms}$, and $\dl_m$
are subalgebras of $\dt_m$, and that $\du_m$ is the contact-analog
of the constant coefficient algebra $\dc_m$.

\begin{lemma} \label{sp iso}
\begin{enumerate} 

\item[(i)]
The space\/ $\dl_m^{\rms}$ is a subalgebra of\/ $\ds_m$ 
isomorphic to\/ $\dsp_{2\ell}$.

\item[(ii)]
The space $\dl_m$ is a Levi subalgebra of\/ $\ds_m$,
with semisimple part\/ $\dl_m^{\rms}$ and center\/ $\bC X_z$.  

\item[(iii)]
The space\/ $\du_m$ is a Heisenberg Lie algebra with center\/ $\bC X_1$.

\item[(iv)]
The affine subalgebra\/ $\dt_m$ is the semidirect sum\/ $\dl_m \oplus_{\rms} \du_m$.
It is a maximal parabolic subalgebra of\/ $\ds_m$ 
with nilradical\/ $\du_m$.

\item[(v)]
The projective subalgebra\/ $\ds_m$ 
is isomorphic to\/ $\dsp_{2(\ell + 1)}$, and
\begin{equation*}
   \ds_m = X \bigl(\theta^{-1} \bigl\{ 
   \mbox{\rm Polynomials of degree~$\le 2$ on $\bR^m$} \bigr\}\bigr).
\end{equation*}

\end{enumerate}
\end{lemma}

\meno{\em Proof.\/}
Most of this can be left to the reader.  
To prove~(i), use~(\ref{Xf}) to obtain
\begin{equation*}
   X_{x_i y_j} = y_j \partial_{y_i} - x_i \partial_{x_j}, \quad
   X_{x_i x_j} = x_i \partial_{y_j} + x_j \partial_{y_i}, \quad
   X_{y_i y_j} = -y_i \partial_{x_j} - y_j \partial_{x_i}.
\end{equation*}
Then note that the natural action of $\dl_m^{\rms}$ 
on $\Span_{\bC} \{ x_i,\th y_i:\th 1 \le i \le \ell \}$ 
preserves the skew-symmetric form defined by 
$\la x_i, x_j \ra = 0$, $\la y_i, y_j \ra = 0$, 
and $\la x_i, y_j \ra = \delta_{ij}$.  
(Alternately, observe that the Lagrange bracket 
defines a non-degenerate $\dl_m$-invariant 
skew-symmetric form on the $-\oh$-eigenspace of $X_z$.)  
Parts~(ii), (iii), and~(iv) now follow by computation.

For a direct proof of the displayed equation in~(v), 
first check that $\E_{xy} = x_i A_i - y_i B_i$.  
Then verify that the vector fields 
$$
A_i,\quad B_i,\quad x_j A_i - y_i B_j,\quad 
x_j B_i - x_i B_j,\quad y_j A_i - y_i A_j,\quad z A_i + \oh y_i \E_{xy},\quad 
z B_i + \oh x_i \E_{xy},
$$ 
$1 \le i, j \le \ell$,
span a $(2\ell + 3) \ell$-dimensional subspace of $\da_m \cap \trm$.

Next, use Lemma~\ref{X} to verify that the right side of the display in~(v)
is a $(2\ell + 3) (\ell + 1)$-dimensional subspace of $\da_m \cap \crm$.  
Since $\da_m$ is $(2\ell + 3) (2\ell + 1)$-dimensional, (v)~is proven.

The fact that $\ds_m$ is a copy of $\dsp_{2(\ell + 1)}$ 
can now be proven using~(ii) and an adaptation of the 
above argument proving that $\da_m$ is a copy of $\dsl_{m+1}$.
$\Box$

\medbreak
Let us establish some notation for later use: we will write $\bC^{2\ell}$
for the basic module of $\dl_m^{\rms} \cong \dsp_{2\ell}$,
and $\Sym^r\bC^{2\ell}$ for its $r^\thup$ symmetric power.
It is well-known that $\Sym^r\bC^{2\ell}$ is self-dual and \irr\ for all~$r$.

The following lemma defines a Cartan subalgebra $\dh_m$ of $\ds_m$
and gives the associated root system and Weyl group.
Its proof may be found in any text on Lie theory;
see \eg\ \cite{Va84}.

\begin{lemma} \label{roots}
\begin{enumerate} 

\item[(i)]
The set\/ $\{ 2X_z,\th X_{x_1 y_1}, \ldots, X_{x_\ell y_\ell} \}$ 
is a basis of a Cartan subalgebra\/ $\dh_m$ of\/ $\ds_m$.  
Under the Cartan-Killing form this basis is orthogonal and 
all of its elements are of length\/ $2 \sqrt{\ell+2}$.  
Let\/ $\{ e_0, e_1, \ldots, e_\ell \}$ 
be the dual basis of\/ $\dh_m^*$.

\item[(ii)]
The adjoint action of\/ $\dh_m$ on 
the polynomial contact vector fields is semisimple:
$X(x^I y^J z^c)$ is of\/ $\dh_m$-weight
  $(2c + |I| + |J| - 2) e_0 - (I_i - J_i) e_i$.

\item[(iii)]
The roots of\/ $\ds_m$ are all of the non-zero\/ $\pm e_i \pm e_j$.  
The Weyl group\/ $W(\ds_m)$ is\/ $S_{\ell+1} \ltimes \bZ_2^{\ell+1}$, 
acting in the usual way on the basis\/ $\{ e_0, \ldots, e_\ell \}$.  

\item[(iv)]
The order\/ $0 < e_\ell < \cdots < e_0$ gives the 
following simple root system\/ $\Pi^+(\ds_m)$ and 
positive root system\/ $\De^+(\ds_m)$ of\/ $\ds_m$:
\begin{equation*}
   \Pi^+(\ds_m) := 
   \bigl\{ e_{i-1} - e_i,\th 2e_\ell \bigr\}_{i=1}^\ell, \quad
   \De^+(\ds_m) := 
   \bigl\{ e_i - e_j \}_{i < j} \cup \{ e_i + e_j \bigr\}_{i,j}.
\end{equation*}
The half-sum of the positive roots is\/ 
$\rho(\ds_m) = \sum_0^\ell (\ell+1-i)e_i$.  
The dominant weights are those\/ $\sum_0^\ell \gamma_i e_i$ 
with\/ $0 \le \gamma_\ell \le \cdots \le \gamma_0$.  
The negative root vectors are\/ $X_1$, $X_{x_i}$, $X_{y_i}$, 
$X_{x_i x_j}$, and those\/ $X_{x_i y_j}$ with\/ $i < j$.

\item[(v)]
The algebra\/ $\dh_m$ is also a Cartan subalgebra of
the Levi subalgebra\/ $\dl_m$ of\/ $\ds_m$.
The roots of\/ $\dl_m$ are the non-zero\/ 
$\pm e_i \pm e_j$ with\/ $i, j > 0$.  It inherits the simple root system
\begin{equation*}
   \Pi^+(\dl_m) = \bigl\{ e_{i-1} - e_i,\th 2e_\ell \bigr\}_{i=2}^\ell.
\end{equation*}

\item[(vi)]
The subalgebra\/ $\dh_m^{\rms}:=\Span\{X_{x_1 y_1}, \ldots, X_{x_\ell y_\ell} \}$
of\/ $\dh_m$ is a Cartan subalgebra of\/ $\dl_m^{\rms}$.
The roots of\/ $\dl_m^{\rms}$ are the same as those of\/ $\dl_m$.

\end{enumerate} 
\end{lemma}

\subsection{The $\ds_m$-structure of the symbol modules} \label{PSA on symbols}

In this subsection we analyze the action of $\ds_m$ 
on the principal and fine symbol modules.  
Our first lemma gives the restriction to $\ds_m$ of
the actions $L^\S_\delta$ and $L^{\FS}_\delta$ 
on $\S_\delta$ and $\FS_\delta$.  
It is a corollary of Proposition~\ref{LXf}.  
In order to state it concisely, 
we define the {\em total weight operator\/} $W$ 
to act on both $\S_\delta$ and $\FS_\delta$ by
\begin{equation*}
   W|_{\S_\delta} = W|_{\FS_\delta} := (\E_z + \oh \E_{xy}) - 
   (\E_\zeta + \oh \E_{\alpha \beta}) + \delta (\ell + 1).
\end{equation*}

\begin{lemma} \label{LXs}

\begin{enumerate}

\item[(i)]
The restrictions of\/ $L^\S_\delta$ and\/ $L^{\FS}_\delta$ 
to the affine subalgebra\/ $\dt_m$ coincide.

\item[(ii)]
Their restriction to the nilradical\/ $\du_m$ of\/ $\dt_m$ 
is the identity map\/ $X_f \mapsto X_f$:
\begin{equation*}
   X_1\ \mapsto\ \pz, \quad 
   X_{x_i}\ \mapsto\ x_i \pz - B_i, \quad
   X_{y_i}\ \mapsto\ y_i \pz - A_i.
\end{equation*}

\item[(iii)]
Their restriction to the Levi factor\/ $\dl_m$ of\/ $\dt_m$ is given by
\begin{eqnarray*}
   X_z &\mapsto& W, \\[6pt]
   X_{x_i y_j} &\mapsto& x_i y_j \pz - (x_i A_j + y_j B_i) + 
   (\alpha_j \partial_{\alpha_i} - \beta_i \partial_{\beta_j}), \\[6pt]
   X_{x_i x_j} &\mapsto& x_i x_j \pz - (x_i B_j + x_j B_i) + 
   (\beta_i \partial_{\alpha_j} + \beta_j \partial_{\alpha_i}), \\[6pt]
   X_{y_i y_j} &\mapsto& y_i y_j \pz - (y_i A_j + y_j A_i) -
   (\alpha_i \partial_{\beta_j} + \alpha_j \partial_{\beta_i}).
\end{eqnarray*}

\item[(iv)]
The action of the rest of\/ $\ds_m$ under\/ $L^{\FS}_\delta$ is given by
\begin{eqnarray*}
  L^{\FS}_\delta (X_{x_i z}) &=& x_i W - z B_i
  - \oh (x_r \alpha_r - y_r \beta_r) \partial_{\alpha_i}
  + \oh \beta_i (x_s \partial_{\beta_s} 
  + y_s \partial_{\alpha_s}), \\[6pt]
  L^{\FS}_\delta (X_{y_i z}) &=& y_i W - z A_i
  + \oh (x_r \alpha_r - y_r \beta_r) \partial_{\beta_i}
  + \oh \alpha_i (x_s \partial_{\beta_s} 
  + y_s \partial_{\alpha_s}), \\[6pt]
  L^{\FS}_\delta (X_{z^2}) &=& 2z W - z^2 \pz - \oh (x_r \alpha_r - y_r \beta_r) 
  (x_s \partial_{\beta_s} + y_s \partial_{\alpha_s}).
\end{eqnarray*}

\item[(v)]
The action of the rest of\/ $\ds_m$ under\/ $L^\S_\delta$ is given by
\begin{eqnarray*}
  L^\S_\delta (X_{x_i z}) &=& L^{\FS}_\delta (X_{x_i z}) + \beta_i \partial_\zeta, \\[6pt]
  L^\S_\delta (X_{y_i z}) &=& L^{\FS}_\delta (X_{y_i z}) + \alpha_i \partial_\zeta, \\[6pt]
  L^\S_\delta (X_{z^2}) &=& L^{\FS}_\delta (X_{z^2}) - (x_r \alpha_r - y_r \beta_r) \partial_\zeta.
\end{eqnarray*}

\end{enumerate}
\end{lemma}

Note that  the operators $x_r \alpha_r - y_r \beta_r$ and 
$x_s \partial_{\beta_s} + y_s \partial_{\alpha_s}$ 
occurring in~(iv) are $\dl^{\rms}_m$-invariants
of total weights~$0$ and~$1$, respectively.

We now define a space which will regularly play an important role
in our arguments:
\begin{equation*}
   (\FS^{k,d}_\delta)^{\du_m} := 
   \bigl\{ P \in \FS^{k,d}_\delta:
   L^{\FS}_\delta(\du_m) P = 0 \bigr\},
\end{equation*}
the subspace of $\FS_\delta^{k,d}$ 
invariant under the Heisenberg algebra $\du_m$.
Since $\dt_m$ normalizes $\du_m$, $(\FS^{k,d}_\delta)^{\du_m}$
is a $\dt_m$-module on which $\du_m$ acts trivially.

\begin{lemma} \label{FS ltrs}
\begin{enumerate}

\item[(i)]
$(\FS^{k,d}_\delta)^{\du_m} = 
\Span_\bC\{ \zeta^{d-k} \alpha^I \beta^J: |I| + |J| = 2k - d \}$.  

\item[(ii)]
Under $W$,\/ $(\FS^{k,d}_\delta)^{\du_m}$ 
has total weight $\delta (\ell+1) -\oh d$.

\item[(iii)]
Under\/ $\dl^{\rms}_m \cong \dsp_{2\ell}$,
$(\FS^{k,d}_\delta)^{\du_m}$ is equivalent to\/ $\Sym^{2k-d} \bC^{2\ell}$.

\end{enumerate}
\end{lemma}

\meno {\em Proof.\/}
Part~(i) is clear from the action of $\du_m$ given in Lemma~\ref{LXs}.
Part~(ii) is straightforward, and Part~(iii) follows from the fact that
as a module of $\dl_m^{\rms}$, $\Span_\bC\{\alpha_i,\beta_i\}_i$
is equivalent to the irreducible module $\bC^{2\ell}$ of $\dsp_{2\ell}$. $\Box$

\medskip

Thus the $\du_m$-invariant fine symbols 
are precisely the constant fine symbols.
In addition to the spaces of constant symbols, we
will encounter the spaces of {\em polynomial fine symbols,\/}
which are modules of the Lie algebra of 
the {\em polynomial contact \vf s.\/}
Under the action of $\ds_m$, these modules turn out
to be the restricted duals of $\dl_m$-relative Verma modules.

To make this precise, let us write
$\Poly(\bR^m)$ for the polynomials in $\cirm$.
We will denote the polynomial subspaces of $\crm$, 
$\F_\lambda(\bR^m)$, $\D_{\lambda, \mu}(\bR^m)$, 
$\S_\delta(\bR^m)$, and $\FS_\delta(\bR^m)$
by writing $\rmp$ in place of $\bR^m$.
Thus for example $\K(\rmp)$ is the classical
Cartan algebra $\K_m$ discussed in the introduction, and
\begin{equation*}
   \FS^{k,d}_\delta (\rmp) := \Span_{\prm} \bigl\{ 
   \zeta^{d-k} \alpha^I \beta^J: |I| + |J| = 2k - d \bigr\}.
\end{equation*}
Note that $\K_m$ is a dense subalgebra of $\crm$ containing $\ds_m$, 
and all of the above polynomial subspaces are dense
$\K_m$-submodules of their smooth counterparts.

We will need the fact that smooth globally defined eigenfunctions
of Euler operators are polynomials.

\begin{lemma} \label{EuLem}
\begin{enumerate}

\item[(i)]
Let $f\in{}C^\infty(\bR^m)$ be an eigenfunction of the full Euler operator 
$\E_u=\sum_{i=1}^m{}u_i\partial_i$ with eigenvalue $\l$.
Then $\l\in\bN$ and $f$ is a homogeneous polynomial in $u$ of degree $\l$.

\item[(ii)]
Let $f\in{}C^\infty(\bR^m)$ be an eigenfunction of the contact Euler operator
$\E_z+\oh\E_{xy}$ with eigenvalue $\l$.
Then $\l\in\oh\bN$ and $f$ is a homogeneous polynomial in 
$(x,y,z)$ of degree $\l$, in the sense that $x_i$ and $y_i$ have degree $\oh$ and $z$
has degree~$1$.

\end{enumerate}
\end{lemma}

\meno {\em Proof.\/}
The first statement is classical, and the second follows from the first
by the change of coordinates $(x,y,z)\mapsto(x,y,z^2)$.
$\Box$

\medskip

Recall from Lemma~\ref{roots} the Cartan subalgebra $\dh_m$ of $\ds_m$.
Given any $\dh_m$-module $V$ and any $\nu \in \dh_m^*$,
we use the standard notation $V_\nu$ for the $\nu$-weight space of $V$.
By the {\em restricted dual\/} of $V$,
we mean the direct sum of the duals of its weight spaces.
A {\em lowest weight vector\/} in an $\ds_m$-module
is a weight vector annihilated by the negative root vectors;
see Lemma~\ref{roots}(iv).
Recall also that total weights are eigenvalues of the operator $W$.

\begin{lemma} \label{FSpolyOne}

\begin{enumerate}

\item[(i)]
$\FS^{k,d}_\delta(\rmp)$ is the span of those elements
of\/ $\FS^{k,d}_\delta$ with well-defined total weight.

\item[(ii)]
$(\FS^{k,d}_\delta)^{\du_m}$ is the lowest 
total weight space of\/ $\FS^{k,d}_\delta(\rmp)$.

\item[(iii)]
$\FS^{k,d}_\delta(\rmp)$ has lowest\/ $\dh_m$-weight
\begin{equation*}
   \nu^{k,d}_\delta = \bigl[ 2\delta (\ell+1) - d \bigr] e_0
   - \bigl[ 2k-d \bigr] e_1.
\end{equation*}
Its lowest\/ $\dh_m$-weight space is the line spanned by 
$\zeta^{d-k} \beta_1^{2k-d}$.

\end{enumerate}
\end{lemma}

\meno {\em Proof.\/}
Lemma~\ref{EuLem} implies~(i), and~(ii) is due to the fact
that polynomials have non-negative total weights.
For~(iii), recall from Lemma~\ref{roots} the basis 
$\{e_0, \ldots, e_\ell\}$ of the weight space $\dh_m^*$
and its order $0 < e_\ell < \cdots < e_0$.
The lowest $\dh_m$-weight space is contained in the lowest
total weight space $(\FS^{k,d}_\delta)^{\du_m}$.
Since the $\dh_m$-weights of $\zeta$, $\alpha_i$,
and $\beta_i$ are, respectively, $-2e_0$, $-e_0 + e_i$, and
$-e_0 - e_i$, (iii)~follows from Lemma~\ref{FS ltrs}.
$\Box$

\medskip
We now recall the basic notions of Verma modules.
Let us denote by $\du^+_m$ and $\dt^+_m$ the subalgebras of $\ds_m$ opposite
to $\du_m$ and $\dt_m$, respectively:
$$
\du^+_m=\Span_\bC\bigl\{X_{x_iz},\,X_{y_iz},\,X_{z^2}\,:\,1\leq{}i\leq\ell\bigr\},
\qquad
\dt_m^+ = \dl_m \oplus_{\rms} \du^+_m.
$$
Given an irreducible module $V$ of $\dl_m$,
the associated relative Verma module
of $\ds_m$ is $\dU(\ds_m) \otimes_{\dt_m^+} V$,
where $\du_m^+$ acts trivially on $V$.

\begin{lemma} \label{FS poly}
As an\/ $\ds_m$-module,\/ $\FS^{k,d}_\delta (\rmp)$ is equivalent to
the restricted dual of the relative Verma module\/
$\dU(\ds_m)\otimes_{\dt_m^+} \Sym^{2k-d} \bC^{2\ell}$.
\end{lemma}

\meno {\em Proof.\/}
By Lemmas~\ref{LXs} and~\ref{FS ltrs}, any non-zero
$\dt_m$-invariant subspace of $\FS^{k,d}_\delta (\rmp)$
contains $(\FS^{k,d}_\delta)^{\du_m}$.
Taking duals, we see that the dual of $(\FS^{k,d}_\delta)^{\du_m}$
generates the restricted dual $\FS^{k,d}_\delta (\rmp)^*$ 
under the action of $\dt_m$.
It follows that $\FS^{k,d}_\delta (\rmp)^*$ is a quotient of 
$\dU(\ds_m)\otimes_{\dt_m^+}
\bigl((\FS_\delta^{k,d})^{\du_m}\bigr)^*$.
The reader may easily check that the two have the same total
weight space dimensions, so they are equivalent.
The result now follows from Lemma~\ref{FS ltrs}(iii)
and the fact that $\bC^{2\ell}$ is self-dual.  $\Box$

\subsection{The $\ds_m$-structure of $\K_m$}

As a particular case of the $\ds_m$-module structures of spaces of fine symbols,
we investigate the $\ds_m$-structure of the algebra $\K_m$ itself, which is
by definition $\FS^{1,2}_0$.  

\begin{prop} \label{K is irr}
The quotient\/ $\K_m/\ds_m$ is \irr\ under\/ $\ds_m$.
Its \lwv\ is $X_{x_1^3}$, which has weight $e_0-3e_1$.
\end{prop}

\meno{\em Proof.\/}
The reader may use Lemma~\ref{LXs} to check that the lowest weight vectors of $\K_m/\ds_m$
under $\dl_m$ are precisely all the elements of the form $X_{x_1^i z^c}$ with $i+c \ge 3$.
The same lemma shows that it is possible to move from any one of these \lwv s to any other
using the elements $X_1$, $X_{x_1}$, $X_{x_1 z}$, and $X_{z^2}$ of $\ds_m$. 
The weight of  $X_{x^3_1}$ is given by Lemma~\ref{roots}(ii). $\Box$

\medskip

It will be important to understand the space of $\du_m$-invariants in $\K_m/\ds_m$.
The following result is immediate from Proposition~\ref{K is irr}.

\begin{cor} \label{WeightK}

\begin{enumerate}

\item[(i)]
$(\K_m/\ds_m)^{\du_m}=\Span\{X_{x^Iy^J}:|I|+|J|=3\}$.

\item[(ii)]
$(\K_m/\ds_m)^{\du_m}$ has total weight $\oh$.
 
\item[(iii)]
$(\K_m/\ds_m)^{\du_m}$ is equivalent to\/ $\Sym^3\bC^{2\ell}$ under the action
of\/ $\dl_m^{\rms}$.

\end{enumerate}

\end{cor}

\subsection{Infinitesimal characters} \label{ICS}

We now turn to the \ic s of the fine symbol modules.
A module of a complex semisimple Lie algebra $\dg$
is said to have an {\em \ic\/} if the
center $\dZ(\dg)$ of the universal enveloping algebra $\dU(\dg)$
acts on it by scalars.  In this case, the \ic\ is
the resulting homomorphism from $\dZ(\dg)$ to $\bC$.
By Schur's lemma, all \irr\ modules have \ic s.

Suppose that a Cartan subalgebra and a positive root system of $\dg$ are fixed.
Let $\rho$ be the half-sum of the positive roots.
If $V$ and $V'$ are two lowest weight modules of $\dg$
with lowest weights $\nu$ and $\nu'$, respectively, and both modules have \ic s,
then it is a consequence of the Harish-Chandra
homomorphism that their \ic s are the same \iff\ $\nu - \rho$
and $\nu' - \rho$ lie in the same orbit of the Weyl group of $\dg$.

\begin{prop} \label{FS eqvs}
Under the action of\/ $\ds_m$, the fine symbol modules have \ic s.  
Fix two fine symbol modules\/ $\FS^{k,d}_\delta$ and\/
$\FS^{k',d'}_\delta$ such that either $k' < k$, or
$k' = k$ and $d' < d$.  They have the same \ic s \iff\ 
at least one of the following four conditions holds:
\begin{equation*} \begin{array}{lrclrcl}
   \mbox{\em (i)\ \ } &
   k' &=& k, &
   d' &=& 2\delta(\ell+1) - 1+2k - d. \\[6pt]
   \mbox{\em (ii)\ \ } &
   k' &=& 2(\delta - 1) (\ell + 1) + 1 - k, &
   d' &=& 2(\delta - 1) (\ell + 1) + 1 - 2k+d. \\[6pt]
   \mbox{\em (iii)\ \ } &
   k' &=& (2\delta-1) (\ell+1) +k- d, &
   d' &=& 2(2\delta-1) (\ell+1) - d. \\[6pt]
   \mbox{\em (iv)\ \ } &
   k' &=& d-k - \ell, &
   d' &=& 2(\delta-1) (\ell+1) +1 - 2k+d.
\end{array} \end{equation*}
Cases~(i), (ii), (iii), and~(iv) cannot occur unless\/
$2\delta (\ell+1)$ is in
\begin{equation*}
   2 + \bN, \quad 2(\ell + 1) + \bN,
   \quad \ell + 2 + \bN, \quad 2\ell + 1 + \bN,
\end{equation*}
respectively.  Moreover, if\/ $\ell = 0$, then Case~(iv)
cannot occur unless $2\delta \in 2 + \bN$.
Therefore if\/ $\delta$ is not contact-resonant (see Section~\ref{Definitions}),
then all of the fine symbol modules of\/ $\D_{\lambda,\mu}$ have distinct \ic s.
\end{prop}
 
\meno {\em Proof.\/}
It is well-known that Verma modules have \ic s,
and so their restricted duals do also.
Therefore by Lemma~\ref{FS poly},
$\dZ(\ds_m)$ acts by scalars on $\FS^{k,d}_\delta(\rmp)$.
By a density argument, it acts by the same scalars on 
$\FS^{k,d}_\delta(\bR^m)$.  
Thus the fine symbol modules have \ic s.

Recall from Lemma~\ref{FSpolyOne} the lowest weight $\nu^{k,d}_\delta$
of $\FS^{k,d}_\delta(\rmp)$, and from Lemma~\ref{roots} the half-sum $\rho(\ds_m)$ 
of the positive roots:
\begin{equation*}
   \rho(\ds_m) = \sum_{i=0}^\ell (\ell+1-i) e_i.
\end{equation*}
Recall also that the Weyl group $W(\ds_m) = S_{\ell+1} \ltimes \bZ_2^{\ell+1}$
acts by permutations and sign changes on the $e_i$.

As stated above, the Harish-Chandra homomorphism shows that
the \ic s of $\FS^{k,d}_\delta$ and $\FS^{k',d'}_\delta$
are the same \iff\ there is an element $w$ of $W(\ds_m)$ such that 
$$
w \bigl(\nu^{k,d}_\delta - \rho(\ds_m)\bigr) = 
\nu^{k',d'}_\delta - \rho(\ds_m).
$$
By Lemma~\ref{FSpolyOne}, such a
$w$ exists \iff\ the two sets
\begin{eqnarray*}
   && \bigl\{ |(2\delta-1)(\ell+1)-d |,\ \ell + 2k - d,\
   \ell - 1,\ \ell - 2,\ \ldots, 2,\ 1 \bigr\}, \\[6pt]
   && \bigl\{  |(2\delta-1)(\ell+1)-d' |,\ \ell + 2k' - d',\
   \ell - 1,\ \ell - 2,\ \ldots, 2,\ 1 \bigr\}
\end{eqnarray*}
are equal.  Since $\ell + 2k - d > \ell - 1 > \cdots > 1 > 0$,
this can occur only in the following ways.

First, $|(2\delta-1)(\ell+1)-d|$ and 
$|(2\delta-1)(\ell+1)-d') |$ can be equal.
In this case $\ell + 2k - d$ and $\ell + 2k' - d'$
must also be equal.  Since we are assuming that
$(k,d)$ and $(k',d')$ are distinct, this leads to Case~(iii).

Second, we could have
\begin{equation*}
   |(2\delta-1)(\ell+1)-d| = \ell + 2k' - d', \qquad
   |(2\delta-1)(\ell+1)-d' | = \ell + 2k - d.
\end{equation*}
This can occur in three ways, depending on the signs 
of the arguments of the absolute values.
If both are negative we arrive at Case~(i),
if both are positive we are in Case~(ii),
and if they are different we obtain Case~(iv).

Finally, if $|(2\delta-1)(\ell+1)-d |$ is equal to
one of $\ell - 1, \ldots, 1$, say~$i$, then again 
$\ell + 2k - d$ and $\ell + 2k' - d'$ must be equal and 
$|(2\delta-1)(\ell+1)-d|$ must also be~$i$.
Therefore here we are still in Case~(iii), albeit with
a different~$w$.  $\Box$

\section{Projective quantization} \label{PQ}

Recall from Section~\ref{SMs} the total symbol modules 
$\S_\delta$ and the total fine symbol modules $\FS_\delta$.
In this section we study quantizations of $\FS_\delta$
invariant under the projective subalgebra $\ds_m$ of $\crm$.
We begin in Section~\ref{S PQ}
with a review of quantizations of $\S_\delta$
invariant under the projective subalgebra $\da_m$ of $\vrm$,
as these quantizations are a component of the
fine projective quantizations.

\subsection{Projective quantization of symbols} \label{S PQ}

A {\em quantization of\/} $\S_\delta$ is defined
to be a linear bijection $Q$ from $\S_\delta$ to 
$\D_{\lambda,\mu}$ which {\em preserves degree\/}
and is the {\em identity on symbols.\/} 
By this we mean that it carries $\S^k_\delta$ into 
$\D^k_{\lambda,\mu}$, and its restriction to $\S^k_\delta$
is a right-inverse of the principal symbol map 
$\sigma^k_{\lambda, \mu}$:
\begin{equation*}
   \sigma^k_{\lambda, \mu} \circ Q|_{\S^k_\delta}:
   \S^k_\delta \to \S^k_\delta \mbox{\rm\ \ is the identity map.}
\end{equation*}
Suppose that $\dg$ is any Lie subalgebra of $\vrm$.
A quantization is said to be a {\em $\dg$-equivariant quantization\/} 
(or simply a {\em $\dg$-quantization})
if it intertwines the two $\dg$-actions 
$L^\S_\delta|_\dg$ and $L_{\lambda,\mu}|_\dg$.

The more vector fields a quantization is invariant 
with respect to, the more useful it is.  
Recall from Section~\ref{SlCoord} the projective subalgebra $\da_m$
and the affine subalgebra $\db_m$ of $\vrm$.
At one extreme, one might ask for a $\vrm$-quantization
of $\S_\delta$.  However, there is no such map: 
$\S_\delta$ and $\D_{\lambda,\mu}$ are not $\vrm$-equivalent
for any $(\lambda, \mu)$.  At the opposite extreme, 
$\db_m$-quantizations are easy to find, 
not unique, and not very useful.
The critical intermediate case is afforded by $\da_m$, because it
is a simple finite dimensional maximal subalgebra of $\vrm$.

\meno {\bf Definition.}
We say that $\delta$ is {\em projectively resonant\/} if it lies in the set 
$$
\left\{1 + \frac{n}{m+1}\;\Big|\; n\in\bN\right\}.
$$ 

\medskip

The following theorem was proven in \cite{CMZ97} for $m=1$,
in \cite{LO99} for arbitrary $m$ at $p=0$, 
and in general in \cite{Le00}.

\begin{thm} \label{PQa}
For $\delta$ not projectively resonant, 
there exists a unique\/ $\da_m$-quantization
\begin{equation*}
   \PQ^{\da_m}_{\lambda,\mu}: \S_\delta \to \D_{\lambda,\mu}.
\end{equation*}
\end{thm}

Theorem~\ref{PQa} may be proven using only the eigenvalues of the Casimir operator;
the full \ic s are not needed.  This is the approach taken in \cite{Le00}.
The explicit formula for the projective quantization
$\PQ^{\da_m}_{\lambda,\mu}$ was given in \cite{CMZ97} for $m=1$,
in \cite{LO99} for arbitrary~$m$ at $p=0$, and in general in \cite{DO01};
see Section~\ref{SlExp} below.

\subsection{Projective quantization of fine symbols} \label{FS PQ}

By analogy with quantizations of $\S_\delta$,
we define a {\em quantization of\/} $\FS_\delta$,
sometimes called a {\em fine quantization,\/} 
to be a linear bijection $Q$ from 
$\FS_\delta$ to $\D_{\lambda,\mu}$,
carrying $\FS^{k,d}_\delta$ into $\D^{k,d}_{\lambda,\mu}$,
which is the identity on fine symbols
in the sense that its restriction
to $\FS^{k,d}_\delta$ is a right-inverse for the fine symbol map
$\fs^{k,d}_{\lambda,\mu}$:
\begin{equation*}
   \fs^{k,d}_{\lambda, \mu} \circ\, Q|_{\FS^{k,d}_\delta}:
   \FS^{k,d}_\delta \to \FS^{k,d}_\delta
   \mbox{\rm\ \ is the identity map.}
\end{equation*}
For any Lie subalgebra $\dg$ of $\crm$,
we say that a fine quantization is a {\em fine $\dg$-equivariant quantization\/}
if it intertwines $L^{\FS}_\delta|_\dg$ and $L_{\lambda,\mu}|_\dg$.

The picture for quantizations of $\FS_\delta$ is 
similar to that for quantizations of $\S_\delta$:
there is no fine $\crm$-quantization, 
there are many fine $\dt_m$-quantizations,
and for most $\delta$ there is a unique fine $\ds_m$-quantization.
The following theorem makes this precise.
It may be proven by combining the results of \cite{DO01} and \cite{FMP08}.
We understand it as a corollary of Proposition~\ref{FS eqvs}.
As remarked before that proposition, it cannot be proven
using the Casimir operator of $\ds_m$ alone; 
the full \ic s of the fine symbol modules are required.

\begin{thm} \label{PQs}
For $\delta$ not contact-resonant (see Section~\ref{Definitions}),
there exists a unique fine\/ $\ds_m$-quantization
\begin{equation*}
   \PQ^{\ds_m}_{\lambda,\mu}: \FS_\delta \to \D_{\lambda,\mu}.
\end{equation*}
\end{thm}

\meno {\em Proof.\/}
Write $\chi^{k,d}_\delta$ for the $\ds_m$-\ic\ of $\FS^{k,d}_\delta$,
and $(\D_{\lambda,\mu})_{\chi^{k,d}_\delta}$ for the subspace
of $\D_{\lambda,\mu}$ on which the center $\dZ(\ds_m)$ of $\dU(\ds_m)$
acts by $\chi^{k,d}_\delta$.  The $\chi^{k,d}_\delta$ are distinct
by Proposition~\ref{FS eqvs}, so we have the $\ds_m$-decomposition
\begin{equation*}
   \D_{\lambda,\mu} =  \bigoplus_{k,d}
   (\D_{\lambda,\mu})_{\chi^{k,d}_\delta}.
\end{equation*}

The fine symbol map $\fs^{k,d}_{\lambda,\mu}$
restricts to the unique fine symbol-preserving $\ds_m$-equivalence
from $(\D_{\lambda,\mu})_{\chi^{k,d}_\delta}$ to $\FS^{k,d}_\delta$.
The fine $\ds_m$-quantization $\PQ^{\ds_m}_{\lambda,\mu}$
is the direct sum of the inverses of these restrictions.  $\Box$

\medskip

The explicit formula for $\PQ^{\ds_m}_{\lambda,\mu}$
will be given in Section~\ref{SpExp}.

\section{Lowest weight calculations} \label{CRCs}

In this section we prove Theorem~\ref{subsymbol} and Theorem~\ref{bifilt}.
The proofs rely on lowest weight calculations in modules of
homomorphisms between the fine symbol spaces.

\subsection{The structure of $\Hom_{\du_m}(\FS^{k,d}_\delta, \FS^{k',d'}_\delta)$}

We begin with a description of the total weight spaces of
$\Hom_{\du_m}(\FS^{k,d}_\delta, \FS^{k',d'}_\delta)$
which will be needed in both proofs.
Recall that for $\nu\in\dh_m^*$, $V_\nu$ denotes the
$\nu$-weight space of any $\dh_m$-module $V$.
We will also use the following notation:
for $w\in\bC$, $V_{(w)}$ denotes the $w$-total weight space of $V$.
Note that if $\nu = \sum_0^\ell \nu_i e_i$,
then $V_\nu \subseteq V_{(\frac{\nu_0}{2})}$.

We will abbreviate $\bC[\pz, A_1, B_1 \ldots, A_\ell, B_\ell]$ by $\bC[\pz, A, B]$.

\begin{lemma} \label{Homu}
\begin{enumerate}

\item[(i)]
The total weight space\/ $\End_{\du_m}\bigl(\cirm\bigr)_{(w)}$ is zero unless\/
$w\in -\oh\bN$, when it is $\bC[\pz, A, B]_{(w)}$.

\item[(ii)]
The total weight space\/ $\Hom_{\du_m}(\FS^{k,d}_\delta, \FS^{k',d'}_\delta)_{(w)}$
is zero unless $w \in \oh(d-d') - \oh \bN$, when it is
\begin{eqnarray*}
   \Span_\bC\bigl\{ \zeta^{d'-k'} \alpha^{I'} \beta^{J'}
   \partial_\zeta^{d-k}\; \partial_\alpha^I\; \partial_\beta^J&:&
   |I|+|J| = 2k-d,\ |I'|+|J'| = 2k'-d' \bigr\} \\[6pt]
   & \otimes & \bC[\partial_z,A,B]_{\left(w-\frac{1}{2}(d-d') \right)}.
\end{eqnarray*}
\end{enumerate}

\end{lemma}

\meno{\em Proof.\/}
Let $T:C^\infty(\bR^m)\to{}C^\infty(\bR^m)$ be of total weight $w$.
Use Lemma~\ref{EuLem} to see that $T$ maps polynomials to polynomials,
and verify that $\End\bigl(\Poly(\bR^m) \bigr)$
is $\bC[x,y,z]\,[[\partial_z,A,B]]$.  Hence
$$
\End\bigl(\Poly(\bR^m)\bigr)_{(w)}=\bC[x,y,z]\,[\partial_z,A,B]_{(w)}.
$$
Since $\partial_z$, $A$, and~$B$ commute with $\du_m$,
the $\du_m$-equivariant maps in this space are precisely those
independent of $x$, $y$, and~$z$.  This proves~(i).

Under $\du_m$, $\FS^{k,d}_\delta$ is equivalent to 
$(\FS^{k,d}_\delta)^{\du_m}\otimes{}C^\infty(\bR^m)$, so
$\Hom_{\du_m}\bigl(\FS^{k,d}_\delta,\FS^{k',d'}_\delta\bigr)$ is
$$
\Hom_{\bC}\bigl((\FS^{k,d}_\delta)^{\du_m},(\FS^{k',d'}_\delta)^{\du_m}\bigr)\otimes
\End_{\du_m} \bigl(C^\infty(\bR^m) \bigr).
$$
Thus~(ii) follows from~(i) and the fact that
$\Hom_{\bC} \bigl((\FS^{k,d}_\delta)^{\du_m},(\FS^{k',d'}_\delta)^{\du_m} \bigr)$
is
$$
  \Span_\bC\bigl\{ \zeta^{d'-k'} \alpha^{I'} \beta^{J'}
   \partial_\zeta^{d-k}\; \partial_\alpha^I\; \partial_\beta^J\;:\;
   |I|+|J| = 2k-d,\ |I'|+|J'| = 2k'-d' \bigr\} .
   \qquad
   \Box
$$

\medskip

We will also need the structure of
$\Hom_{\du_m} \bigl(\FS^{k,d}_\delta, \FS^{k',d'}_\delta \bigr)_{(w)}$
as a module of the subalgebra $\dl_m^{\rms}$,
which recall is isomorphic to $\dsp_{2\ell}$.

\begin{lemma} \label{SymLem}
Suppose that $w$ is in $\oh(d-d') - \oh \bN$.
Then under the action of\/ $\dl_m^{\rms}$,
$$
\Hom_{\du_m} \bigl(\FS^{k,d}_\delta, \FS^{k',d'}_\delta \bigr)_{(w)}
\th \cong \th
\Sym^{2k-d}\bC^{2\ell}\,\otimes\,\Sym^{2k'-d'}\bC^{2\ell}\,\otimes\,
\biggl(\th
\bigoplus_{r\geq0}\Sym^{d-d'-2r-2w}\bC^{2\ell}
\biggr),
$$
where the direct sum is only over non-negative exponents.
\end{lemma}

\meno{\em Proof.\/}
Consider the explicit basis of
$\Hom_{\du_m}\bigl(\FS^{k,d}_\delta, \FS^{k',d'}_\delta\bigr)_{(w)}$
given in Lemma~\ref{Homu}(ii).  We have the following $\dl_m^{\rms}$-equivalences:
$$
\begin{array}{rcl}
\Span_\bC
\bigl\{ \alpha^{I'} \beta^{J'}\;:\;
|I'|+|J'| = 2k'-d' \bigr\} &\cong& \Sym^{2k'-d'}\bC^{2\ell},\\[6pt]
\Span_\bC\bigl\{
\partial_\alpha^I\; \partial_\beta^J\;:\;
|I|+|J| = 2k-d\bigr\}  &\cong& \Sym^{2k-d}\bC^{2\ell},
\end{array}
$$
and for $v \in -\oh\bN$, $\bC[\partial_z,A,B]_{(v)} \cong
\bigoplus_{r\geq0}\Sym^{-2v-2r}\bC^{2\ell}$.
Since $\zeta$, $\partial_\zeta$, and $\partial_z$ are $\dl_m^{\rms}$-invariant,
the result follows.
$\Box$

\subsection{Proof of Theorem~\ref{subsymbol}}

Assume temporarily that $M$ is $\bR^m$, equipped with the standard contact structure.
Since $\delta$ is not contact-resonant,
Theorem~\ref{PQs} shows that there is a unique $\ds_m$-equivariant
map from $\D^k_{\l,\mu}$ to $\FS^{k-1,\,2(k-1)}_\delta$, namely
\begin{equation} \label{ssEQ}
\ss^k_{\l,\mu} := \pi_{k-1,\,2(k-1)} \circ (\PQ^{\ds_m}_{\lambda,\mu})^{-1},
\end{equation}
where $\pi_{j,d}$ denotes the canonical projection from
$\FS_\delta$ to $\FS^{j,d}_\delta$.
Our task is to prove that this map is in fact $\crm$-equivariant.  

We use the projective quantization to pull the action $L_{\lambda,\mu}$ of
$\crm$ on $\D_{\l,\mu}$ back to an action $\L_{\lambda, \mu}$ of $\crm$ on $\FS_\delta$: 
\begin{equation*}
   \L_{\l,\mu}(X_f)\ :=\ (\PQ^{\ds_m}_{\lambda,\mu})^{-1} \circ
   L_{\l,\mu}(X_f) \circ \PQ^{\ds_m}_{\lambda,\mu}.
\end{equation*}
The statement that $\ss^k_{\l,\mu}$ is $\crm$-equivariant is equivalent
to the statement that
$$
\pi_{k-1,\,2(k-1)}: \bigoplus_{0\leq{}j\leq{}k} \bigoplus_{j\leq{}d\leq{}2j}
\FS^{j,d}_\delta \to\FS^{k-1,\,2(k-1)}_\delta
$$ 
intertwines the $\crm$-actions $ \L_{\l,\mu}$ and $L^{\FS}_\delta$.

We may regard $ \L_{\l,\mu}$ as a block matrix with entries
\begin{equation*}
   \L_{\l,\mu}^{(j,d),(j',d')}(X_f):
   \FS^{j,d}_\delta \to \FS^{j',d'}_\delta. 
\end{equation*}
This matrix is triangular with respect to the dictionary order on $(j,d)$,
and the diagonal entries are simply the usual actions on fine symbols:
$\L_{\l,\mu}^{(j,d),(j,d)}$ is $L^{\FS}_\delta$ restricted to $\FS^{j,d}_\delta$.
As a result of the $\ds_m$-equivariance of $\PQ^{\ds_m}_{\l,\mu}$,
the off-diagonal entries $\L_{\l,\mu}^{(j,d),(j',d')}$ are $\ds_m$-relative
1-cochains of $\crm$: they vanish on $\ds_m$ and are $\ds_m$-equivariant maps from
$\crm/\ds_m$ to $\Hom(\FS^{j,d}_\delta, \FS^{j',d'}_\delta)$.

One finds that $\pi_{k-1,\,2(k-1)}\circ{}\L_{\l,\mu}$
restricted to $\bigoplus_{0\leq{}j\leq{}k}\bigoplus_{j\leq{}d\leq{}2j}\FS^{j,d}_\delta$ is
$$
L_\delta^{\FS}\circ\pi_{k-1,\,2(k-1)}+
\sum_{d=k}^{2k}\L_{\l,\mu}^{(k,d),(k-1,\,2(k-1))}\circ\pi_{k,d}.
$$
Therefore it suffices to prove that entries
$\L_{\l,\mu}^{(k,d),(k-1,\,2(k-1))}$ are zero for all $d$.
The following lemma is the key to the situation.

\begin{lemma} \label{NoExist}
Let\/ $V$ be a module of\/ $\crm$ such that 
the space\/ $V_{(\frac{1}{2})}^{\du_m}$
of\/ $\du_m$-invariants in\/ $V$ of total weight\/ $\oh$
contains no copies of\/ $\Sym^3\bC^{2\ell}$ under the action of\/ $\dl_m^{\rms}$.
Then the space\/ $C^1\bigl(\K(\bR^m), \ds_m; V\bigr)$ of\/
$\ds_m$-relative 1-cochains of\/ $\crm$ with values in\/ $V$ is zero.
\end{lemma}

\meno {\em Proof.\/}
Apply Corollary~\ref{WeightK}. $\Box$

\medskip

In order to apply this lemma, we must prove that
$\Hom_{\du_m} \bigl(\FS^{k,d}_\delta, \FS^{k-1,\,2(k-1)}_\delta \bigr)_{(\frac{1}{2})}$
contains no copies of\/ $\Sym^3\bC^{2\ell}$ under $\dl_m^{\rms}$
for $k \le d \le 2k$.  When $d=2k$, by Lemma~\ref{Homu} we obtain
$$
\Hom_{\du_m}\bigl(\FS^{k,2k}_\delta, \FS^{k-1,\,2(k-1)}_\delta\bigr)_{(\frac{1}{2})}=
\zeta^{k-1}\partial_\zeta^k\,\Span\{A,B\},
$$
which is a copy of\/ $\bC^{2\ell}$.
When $d=2k-1$, we obtain
$$
\Hom_{\du_m} \bigl(\FS^{k,2k-1}_\delta, \FS^{k-1,\,2(k-1)}_\delta \bigr)_{(\frac{1}{2})}=
\zeta^{k-1}\partial_\zeta^{k-1}\,\Span\{\partial_\alpha,\partial_\beta\},
$$
which is again a copy of $\bC^{2\ell}$.
Finally, when $d<2k-1$, we obtain zero.

This completes the proof of Theorem~\ref{subsymbol} when $M=\bR^m$.
If $M$ is an arbitrary contact manifold,
Darboux's theorem implies the existence of an atlas of local charts on $M$
that are diffeomorphic to the standard contact structure on $\bR^m$.
We have just proved that for every chart $U$, there is a unique (locally defined) map
$\ss^k_{\l,\mu}(U)$ equivariant with respect to $\K(U)$.
If $U'$ is another chart, uniqueness implies that
the maps $\ss^k_{\l,\mu}(U)$ and $\ss^k_{\l,\mu}(U')$ coincide on $U\cap{}U'$.
Therefore $\ss^k_{\l,\mu}(M)$ is well-defined on all of $M$ and obviously commutes with $\K(M)$.

\subsection{Proof of Theorem~\ref{bifilt}} \label{C proof}

As in the proof of Theorem~\ref{subsymbol}, take $M=\bR^m$
and consider the action $\L_{\lambda,\mu}$ of\/ $\crm$ on $\FS_\delta$.
We first prove that certain of its matrix elements
$\L_{\l,\mu}^{(k,d),(k',d')}$ vanish.

\begin{lemma} \label{k-1}
In both of the following cases, the space of\/ $\ds_m$-relative 1-cochains
of\/ $\crm$ with coefficients in\/
$\Hom \bigl(\FS^{k,d}_\delta, \FS^{k',d'}_\delta \bigr)$ is zero:

\begin{enumerate}

\item[(i)]
$k' = k-1$ and $d' \ge d$.

\item[(ii)]
$k' < k-1$ and $d' \ge d - (k-k') + 2$.

\end{enumerate}
\end{lemma}

\meno{\em Proof.\/}
By Lemma~\ref{NoExist}, we must prove that under $\dl_m^{\rms}$ the space
$\Hom_{\du_m}(\FS^{k,d}_\delta, \FS^{k',d'}_\delta)_{(\frac{1}{2})}$
contains no copies of $\Sym^3\bC^{2\ell}$ in either Case~(i) or Case~(ii).
By Lemma~\ref{Homu}, this space is zero unless $d-d' \in \bZ^+$.
In particular, it is zero in Case~(i).
In Case~(ii) with $d-d' \in \bZ^+$, Lemma~\ref{SymLem}
shows that under $\dl_m^{\rms}$ it is equivalent to
\begin{equation} \label{PRV}
   \Sym^{2k-d} \bC^{2\ell}\, \otimes\,\Sym^{2k'-d'} \bC^{2\ell}\, \otimes\,
   \biggl(\th \bigoplus_{r\geq0}\Sym^{d-d'-2r-1}\bC^{2\ell} \biggr).
\end{equation}

It is well-known (see, \eg\ \cite{Va84}) that the largest irreducible component of
$\Sym^r\bC^{2\ell} \otimes{} \Sym^s\bC^{2\ell}$
is $\Sym^{r+s}\bC^{2\ell}$, and for $r\geq{}s$, its smallest
irreducible component is its Parthasarathy - Ranga Rao - Varadarajan submodule,
a copy of $\Sym^{r-s}\bC^{2\ell}$.
Therefore the maximal component of 
$$
\Sym^{2k'-d'}\bC^{2\ell}\,\otimes\,
\biggl(\th \bigoplus_{r\geq0}\Sym^{d-d'-2r-1}\bC^{2\ell} \biggr)
$$ 
is $\Sym^{d-2d'+2k'-1}\bC^{2\ell}$.
In Case~(ii) we have $k'-d'<k-d$, so the smallest
irreducible component of~(\ref{PRV}) is $\Sym^{2(d'-d+k-k')+1}\bC^{2\ell}$,
which is larger than $\Sym^{3}\bC^{2\ell}$.
$\Box$

\medskip

This lemma implies that $\L_{\l,\mu}^{(k,d),(k',d')}=0$
under the conditions of Cases~(i) and~(ii).
Therefore the space $\D^{(b)}_{\l,\mu}(\bR^m)$ defined by
\begin{equation*}
\D^{(b)}_{\l,\mu}(\bR^m) \th:=\th
\PQ^{\ds_m}_{\l,\mu} \biggl(\th \bigoplus_{2d-k\leq{}b}\FS^{k,d}_\delta \biggr)
\end{equation*}
is invariant under the action $L_{\l,\mu}$ of $\K(\bR^m)$.
Since under $\ds_m$ there is unique copy of $\FS^{k,d}_\delta$ in $\D_{\l,\mu}$
for all $(k,d)$, $\D^{(b)}_{\l,\mu}$ is the unique subspace of $\D_{\l,\mu}$
whose graded module is as in Theorem~\ref{bifilt}.
This completes the proof for $M = \bR^m$.
For $M$ arbitrary, local existence and uniqueness allows us to conclude
global existence and uniqueness of $\D^{(b)}_{\l,\mu}(M)$
as in the proof of Theorem~\ref{subsymbol}.

\section{Explicit formulas} \label{T}

\subsection{The affine invariants}

We begin by proving that the following maps between fine symbol modules
are equivariant with respect to the affine subalgebra $\dt_m$:

\meno{\bf Definition.}
The {\em contact divergence\/} is the map 
\begin{equation*}
   \DivC := \pz \partial_\zeta: 
   \FS^{k, d}_\delta \to \FS^{k-1, d-2}_\delta.
\end{equation*}
The {\em tangential divergence\/} is the map
\begin{equation*}
   \DivT := A_r \partial_{\alpha_r} + B_r \partial_{\beta_r}:
   \FS^{k, d}_\delta \to \FS^{k-1, d-1}_\delta.
\end{equation*}
Finally, define
\begin{equation*}
   \De := (\alpha_r B_r - \beta_r A_r) \partial_\zeta:
   \FS^{k, d}_\delta \to \FS^{k, d-1}_\delta.
\end{equation*}

\begin{lemma} \label{divergences}
\begin{enumerate}

\item[(i)]
$\DivC$, $\DivT$, and\/ $\De$ are all $\dt_m$-equivariant.

\item[(ii)]
The contact divergence\/ $\DivC$ commutes 
with both\/ $\DivT$ and\/ $\De$, and
\begin{equation*}
   [\DivT, \De] = (\ell + \E_{\alpha \beta}) \DivC.
\end{equation*}

\item[(iii)]
Regarded as a map from\/ $\S^k_\delta$ to\/ $\S^{k-1}_\delta$,\/
$\DivT + \DivC$ is the full divergence\/ $\Div$.
\end{enumerate}
\end{lemma}

\meno {\it Proof.}
The statement follows from Lemma~\ref{LXs} and short computations.  
$\Box$

\meno {\bf Remark.}
It follows from the first fundamental theorem of invariant theory
for $\dsp_{2\ell}$ that the associative algebra generated by the operators
$$
\DivC,
\qquad
\DivT,
\qquad
\De,
\qquad
\E_{\alpha \beta},
\qquad
\E_{\zeta}
$$
coincides with the algebra $\End_{\dt_m}(\FS_\delta)$
of all affine invariants: see \cite{FMP07}.

\subsection{The $\dsl_{m+1}$-equivariant quantization of $\S_\delta$} \label{SlExp}

Continuing the discussion of Section~\ref{S PQ}, we now give an explicit formula
for the quantization map $\PQ^{\da_m}_{\lambda,\mu}$ of Theorem~\ref{PQa}.
This formula is part of the explicit formula 
for the fine projective quantization in the contact setting.

We begin with the standard symbol calculus on $\S_\delta$.
As in Section~\ref{SlCoord},
fix any coordinates $u_1, \ldots, u_m$ on $\bR^m$.
Write $\xi_i$ for the symbol of the \vf\ $\partial_{u_i}$.
Then
\begin{equation*}
   \S_\delta = \Span_{\cirm} \bigl\{ \xi^I: I \in \bN^m \bigr\},
   \qquad
   \D_{\lambda,\mu} = \Span_{\cirm} \bigl\{ \partial_u^I: I \in \bN^m \bigr\}.
\end{equation*}
As noted in Section~\ref{SMs}, these spaces carry the $\vrm$-actions
$L^\S_\delta$ and $L_{\lambda,\mu}$, respectively.

One of the simplest quantizations is
the {\em normal order quantization:\/} 
\begin{equation*}
   \NQ: \S_\delta \to \D_{\lambda,\mu}, \qquad
   \NQ \bigl( f_I(u) \xi^I \bigr)
   := f_I(u) \partial_u^I.
\end{equation*}
In the literature, $\S_\delta$ and $\D_{\lambda,\mu}$
are frequently identified via $\NQ$, which then
does not appear explicitly in the formulas.
Although $\NQ$ is not an $\da_m$-equivariant quantization,
it does turn out to be a $\db_m$-quantization.
The first step in computing $\PQ^{\da_m}_{\lambda,\mu}$
is to find explicit formulas
for $L^\S_\delta$ and $L_{\lambda,\mu}$.
In fact, one computes the pull-back of $L_{\lambda,\mu}$
to $\S_\delta$ via $\NQ$, that is,
\begin{equation*}
   L_{\lambda,\mu}^{\NQ^{-1}} (X)\ :=\
   \NQ^{-1} \circ L_{\lambda,\mu}(X) \circ \NQ.
\end{equation*}
Towards this end, note that any \dog\ on $\S_\delta$ may be written as a
$\cirm$-linear combination of monomials $\partial_u^I \xi^J \partial_\xi^K$.

Given any vector field $X = \sum_1^m X_i \partial_{u_i}$,
it is straightforward to obtain
\begin{equation*} 
   L_{\lambda,\mu}^{\NQ^{-1}} (X) = X + \delta \Div(X)
   - \sum_{|I| > 0} \ts\frac{1}{I!} 
   \Bigl( \partial_u^I \bigl[ \lambda \Div(X)
   + {\ts\sum_j X_j \xi_j} \bigr]\Bigr) \partial_\xi^I.
\end{equation*}
The action $L^\S_\delta$ is simply the part of 
$L_{\lambda,\mu}^{\NQ^{-1}}$ which preserves $\xi$-degree:
\begin{equation*} 
   L^\S_\delta(X) = X + \delta \Div(X)
   - \sum_{i,j=1}^m (\partial_{u_i} X_j) \xi_j \partial_{\xi_i}.
\end{equation*}
Observe that these two formulas are the same \iff\
$X \in \db_m$.  Thus as claimed, $\NQ$ is a $\db_m$-quantization.

As usual, let $\E_\xi$ denote the $\xi$-Euler operator.
The full divergence operator is
\begin{equation*}
   \Div = \sum _{1\leq{}i\leq{}m} \partial_{u_i} \partial_{\xi_i}
   : \S^k_\delta \to \S^{k-1}_\delta.
\end{equation*}
Bear in mind that $\E_\xi$ and $\Div$ do not commute:
$[\E_\xi, \Div] = -\Div$.  The theorem is as follows.

\begin{thm} \label{PQa formula} \cite{LO99, DO01}  
For $\delta$ non-resonant,
\begin{equation*}
   \PQ^{\da_m}_{\lambda,\mu} = \NQ \circ 
   \sum_{s=0}^\infty {\ts \frac{1}{s!}} \Div^s 
   \circ {\E_\xi + \lambda(m+1) - 1 \choose s}
   {2\E_\xi - \delta(m+1) + m - 1 \choose s}^{-1}.
\end{equation*}
\end{thm}

It is worth mentioning that this formula can be
understood as a (non-commutative) hypergeometric function: see \cite{DO01}.

\subsection{The $\dsp_{2(\ell+1)}$-equivariant quantization of $\FS_\delta$} \label{SpExp}
We now proceed to derive an explicit formula for $\PQ^{\ds_m}_{\lambda,\mu}$.
We begin by defining a map $\SPQ^{\ds_m}_\delta$ from $\FS_\delta$ to $\S_\delta$:
\begin{equation} \label{SPQ formula}
   \SPQ^{\ds_m}_\delta := \sum_{s=0}^\infty { \frac{2^s}{(s!)^2}}
   \Delta^s \circ {2\E_\zeta - 2\delta (\ell + 1) \choose s}^{-1}.
\end{equation}
This formula is well-defined
provided that $\delta$ is not contact-resonant.
Recall from~(\ref{symbol bases}) that we are abusing notation
and using the same bases for $\FS_\delta$ and $\S_\delta$
(this is analogous to regarding the normal order quantization
$\NQ$ as the identity).
Therefore we may and do regard $\SPQ^{\ds_m}_\delta$
as a map from $\FS_\delta$ to $\S_\delta$.

\begin{thm} \label{SPThExp} 
The fine\/ $\ds_m$-equivariant quantization of Theorem~\ref{PQs} is 
$$
\PQ_{\l,\mu}^{\ds_m}=\PQ_{\l,\mu}^{\da_m}\circ\SPQ^{\ds_m}_\delta.
$$
\end{thm}

\meno{\it Proof.\/}
Observe that it suffices to prove that $\SPQ^{\ds_m}_\delta$
intertwines the restrictions of the actions
$L^{\FS}_\delta$ and $L^\S_\delta$ to $\ds_m$.
By Lemma~\ref{divergences}, $\SPQ^{\ds_m}_\delta$ intertwines the $\dt_m$-actions.
Restricted to $\FS^{k,d}_\delta$, it is of the form $\sum_0^\infty \Delta^s C_s$
for some constants $C_s$.  One obtains~(\ref{SPQ formula}) by deriving
a recursion relation for these constants.
Since $X_{x_i z}$ generates $\ds_m$ under $\dt_m$ for any $i$,
we need only impose the condition
$$
L^\S_\delta (X_{x_i z}) \circ \SPQ^{\ds_m}_\delta = 
\SPQ^{\ds_m}_\delta \circ L^{\FS}_\delta (X_{x_i z}).
$$

By Lemma~\ref{LXs}, $L^\S_\delta (X_{x_i z}) - L^{\FS}_\delta(X_{x_i z})$
is $\beta_i \partial_\zeta$, which commutes with $\Delta$.
Therefore we find that $\SPQ^{\ds_m}_\delta\circ \beta_i \partial_\zeta$
and $\bigl[\SPQ^{\ds_m}_\delta, L^{\FS}_\delta(X_{x_i z}) \bigr]$ must be equal, \ie\
\begin{equation*}
   0 = \ts \sum_s \bigl( \Delta^s \circ \beta_i \partial_\zeta 
   - \bigl[\De^s, L^{\FS}_\delta(X_{x_i z}) \bigr] \bigr) C_s.
\end{equation*}
Using the same lemma, deduce the following commutator:
$$
   [L^{\FS}_\delta(X_{x_i z}), \Delta^s]= -s \Delta^{s-1} \beta_i \partial_\zeta
   \bigl( \E_\zeta - \delta(\ell+1) - \oh(s-1) \bigr),
$$
This gives the recursion relation
$$
   C_{s-1} = \oh s \bigl( 2c - 2\delta (\ell+1) - (s-1) \bigr) C_s.
$$
Since $C_0 = 1$, the theorem follows.  $\Box$

\medbreak

\subsection{The subsymbol} \label{SubsExp}

In order to give an explicit local formula for $\ss^k_{\l,\mu}$,
let us fix a system of Darboux coordinates as in Section~\ref{DarbSect}.
An arbitrary differential operator $T$ of order $\leq{}k$ may be expressed as
$$
T=\sum_{c+|I|+|J|\leq{}k}T_{c,I,J}\,\partial_z^c\partial_x^I\partial_y^J,
$$
where the $T_{c,I,J}$ are smooth functions.  As usual, we
replace $\partial_z,\partial_x,\partial_y$ by their symbols
$\xi_z,\xi_x,\xi_y$, respectively.  This amounts to replacing $T$ by
$$
\NQ^{-1}(T)=
\sum_{c+|I|+|J|\leq{}k}T_{c,I,J}\,\xi_z^c\xi_x^I\xi_y^J.
$$

The formula has two ingredients: the full divergence $\Div$,
and the projection $\pi_{k-1,2(k-1)}:\S^{k-1}_\delta\to\FS^{k-1,2(k-1)}_\delta$.
The full divergence $\Div(\NQ^{-1}(T))$ is 
$$
\sum_{c+|I|+|J|\leq{}k}\biggl(
c\partial_z(T_{c,I,J})\,\xi_z^{c-1}\xi_x^I\xi_y^J+
\sum_{s=1}^\ell{}\Bigl(
I_s\partial_{x_s}(T_{c,I,J})\,\xi_z^c\xi_x^{I-e_s}\xi_y^J+
J_s\partial_{y_s}(T_{c,I,J})\,\xi_z^c\xi_x^{I}\xi_y^{J-e_s}
\Bigr)\biggr).
$$
In the $(\alpha, \beta, \zeta)$-coordinates on $\S^{k-1}_\delta$
given in Section~\ref{SMs}, the projection $\pi_{k-1,2(k-1)}$ simply gives the $\zeta^{k-1}$ term.
In Darboux coordinates,
$$
\pi_{k-1,2(k-1)}\bigl(T_{c,I,J}\,\xi_z^{k-1-|I|-|J|}\xi_x^I\xi_y^J\bigr)=
(-1)^{|I|}\bigl(\oh\bigr)^{|I|+|J|}y^{I}x^{J}
T_{c,I,J}\zeta^{k-1},
$$
because by~(\ref{abxi}), $\xi_{x_i}=\alpha_i-\oh{}y_i\zeta$, 
$\xi_{y_i}=-\beta_i+\oh{}x_i\zeta$, and $\xi_z=\zeta$.

\begin{prop} \label{subsymbol formula}
In Darboux coordinates, the subsymbol\/ $\ss^k_{\l,\mu}(T)$ is given by
$$
\ss^k_{\l,\mu}(T) =\pi_{k-1,2(k-1)}\circ\Bigl(
1-\frac{(k-1)+2\l(\ell+1)}{2(k-1)-2(\delta-1)(\ell+1)}\Div
\Bigr)\circ{}\NQ^{-1}(T).
$$
\end{prop}

\meno{\it Proof.\/}
Using~(\ref{ssEQ}) and Theorem~\ref{SPThExp}, we obtain
$\ss^k_{\l,\mu}(T) =\pi_{k-1,2(k-1)}\circ
\bigl(\SPQ^{\ds_m}_\delta\bigr)^{-1}
\circ
\bigl(\PQ_{\l, \mu}^{\da_m}\bigr)^{-1}$.
Since $\FS^{k-1,2(k-1)}_\delta$ is the fine symbol module in $\S^{k-1}_\delta$
of the highest contact order, (\ref{SPQ formula}) gives
$$
\pi_{k-1,2(k-1)}\circ
\bigl(\SPQ^{\ds_m}_\delta\bigr)^{-1}=\pi_{k-1,2(k-1)}.
$$
To calculate the $\S^{k-1}_\delta$-component of $\bigl(\PQ_{\l, \mu}^{\da_m}\bigr)^{-1}$,
we use the formula of Theorem~\ref{PQa formula}.
$\Box$

\medskip

Note that the formula for $\ss^k_{\l,\mu}$ is well-defined for
all but one contact-resonant value of $\delta$, the value
$\delta=\frac{\ell+k}{\ell+1}$.
By continuity, it retains $\K_m$-equivariance whenever it is well defined.

\subsection{Proof of Theorem~\ref{subsOrd2}} \label{thmB}

Let $T$ be a second order operator from $\F_\l$ to $\F_\l$.
Since $\FS^{1,2}_0$ is equivalent to $\F_{-\frac{1}{\ell+1}}$,
the subsymbol $\ss^2_{\l,\l}(T)$ may be written as a contact Hamiltonian.
If the operator $T$ is of the form
$$
\begin{array}{rcl}
T&=&T_{2,0,0}\,\partial_z^2+T_{1,i,0}\,\partial_z\partial_{x_i}
+T_{1,0,i}\,\partial_z\partial_{y_i}+T_{0,ij,0}\,\partial_{x_i}\partial_{x_j}+
T_{0,i,j}\,\partial_{x_i}\partial_{y_j}+T_{0,0,ij}\,\partial_{y_i}\partial_{y_j}\\[4pt]
&&+T_{1,0,0}\,\partial_z+T_{0,i,0}\,\partial_{x_i}+T_{0,0,i}\,\partial_{y_i}+T_{0,0,0},
\end{array}
$$
then the formula of Proposition~\ref{subsymbol formula} reads
$$
\begin{array}{rcl}
\ss^2_{\l,\l}(T) &=&
-\frac{1+2\l(\ell+1)}{\ell+2}
\Bigl(
\partial_z(T_{2,0,0}-\oh{}y_iT_{1,i,0}+\oh{}x_iT_{1,0,i})\\[6pt]
&&\hskip1.6cm
+\partial_{x_i}(T_{1,i,0}-\oh{}y_jT_{0,ij,0}+\oh{}x_jT_{0,i,j})\\[6pt]
&&\hskip1.6cm
+\partial_{y_i}(T_{1,0,i}+\oh{}x_jT_{0,0,ij}-\oh{}y_jT_{0,j,i})\Bigr)\\[6pt]
&&
+T_{1,0,0}-\oh{}y_iT_{0,i,0}+\oh{}x_iT_{0,0,i}.
\end{array}
$$
One can verify directly that this formula coincides with that of Theorem~\ref{subsOrd2}.
The most efficient approach is to prove first the comments following the statement of the theorem,
after which it suffices to carry out the verification for the single operator
$T = L_\lambda(X_z) \circ L_\lambda(A_1)$, because that determines $c_{13}$.

\medskip

\noindent \textbf{Acknowledgments}.
The first author was partially supported by 
Simons Foundation Collaboration Grant 207736.
The second author was partially supported by Grant
PICS05974, ``PENTAFRIZ'', of the CNRS.
The article was begun while both authors were supported by the
Research in Pairs program of the
Mathematisches Forschungsinstitut Oberwolfach,
for whose support they are very grateful.
They are also pleased to thank
Dmitry Fuchs, J\'er\^ome Germoni, Peter Littelmann,
Sophie Morier-Genoud, Sergei Tabachnikov, and
Andr\'e Unterberger for enlightening discussions.

\def\eightit{\it} 
\def\bib{\bf}
\bibliographystyle{amsalpha}

\end{document}